\documentclass[10pt,pdftex]{iopart}
\usepackage{iopams}

\usepackage{tensind,graphicx,gensymb,bbm,wasysym,amssymb,gensymb,bbm}
\usepackage{mathptmx}      
\usepackage{color}
\usepackage{units}
\usepackage{url}

\begin{document}
\title{Numerical Chladni figures}

\author{Thomas M{\"u}ller}
\address{
  Visualisierungsinstitut der Universit\"at Stuttgart (VISUS)\\
  Allmandring 19, 70569 Stuttgart, Germany
}
\ead{Thomas.Mueller@visus.uni-stuttgart.de}

\begin{abstract}
Chladni patterns of vibrating membranes or thin plates faszinated the people already in the eighteenth century. As a simple way to visualize acoustic phenomena it is a valuable experiment for beginners' courses. In this paper I present \emph{NumChladni}, an interactive tool for studying arbitrary two-dimensional vibrating membranes based on the Finite Element method. I also describe the straightforward approach of the underlying mathematical details and give some examples. \emph{NumChladni} is directly applicable in the undergraduate classroom as, for example, complementary application to experimental setups.
\end{abstract}

%







\section{Introduction}\label{sec:intro}
The eigenmodes of a two-dimensional vibrating membrane or thin plate can be visualized using fine sand. However, not the eigenmodes themself but only the nodal lines become visible where the sand accumulates. These nodal line patterns where first published by Chladni~\cite{chladni1787} in 1787 who attracted the public's attention for these also esthetic figures. 

While the eigenmodes of a rectangular or a circular membrane can be given in closed form by solving the wave equation analytically, the eigenmodes of arbitrarily shaped membranes are much more difficult to determine. 
A straightforward numerical approach is to use the finite element (FE) method where the membrane is approximated for example by a triangular mesh. The eigenmodes then follow from solving an eigenvalue problem for the mesh vertices. Depending on the mesh size, the quality of the solution can be improved but the numerical effort increases considerably. As long as we restrict to low frequencies, the finite element method delivers quite satisfactory eigenmodes.

So far, Chladni patterns where discussed in the educational literature only from an experimental point of view. Rossing~\cite{rossing:271}, for example, gave some hints on how to best excite plate vibrations and how to observe the patterns. Comer et al.~\cite{comer:1345} demonstrated that dust could also collect at the antinodes (see also the comment by Rossing~\cite{rossing:283} on the Comer article). A more detailed historical outline as well as an extensive collection of Chladni figures can be found in the book by Waller~\cite{waller}.

The aim of this article is to give a concise introduction to the computational study of the eigenmodes of arbitrarily shaped membranes by means of the finite element method which could be used as introductory material for courses in elasticity theory or acoustics. 
Although this straightforward approach yields only the lowest frequencies with an acceptable accuracy, it nevertheless gives a valid impression of the structure of the eigenmodes.
Additionally, \emph{NumChladni} is presented which is a graphical user interface to interactively explore the eigenmodes calculated by the finite element method. It is directly applicable in the undergraduate classroom as complementary application to experimental studies, for example.

The structure of the paper is as follows. In Sec.~\ref{sec:membranes} there is a step-by-step derivation, strongly based on the book by Schwarz~\cite{schwarzFEM}, on how the eigenmodes of a membrane can be determined by means of the finite element method. The graphical user interface \emph{NumChladni} as well as the necessary libraries to generate the finite element mesh and to solve the corresponding eigenvalue system are presented in Sec.~\ref{sec:numchladni}. A few examples and possible exercises are given in Sec.~\ref{sec:examples}.

\emph{NumChladni} is freely available for Linux and Windows. The source code and several examples can be downloaded from 
\url{http://go.visus.uni-stuttgart.de/numchladni}.

%
\section{Vibrating membranes}\label{sec:membranes}
The eigenmodes of a two-dimensional vibrating membrane follow from the Bernoulli solution of the wave equation
\begin{equation}
  \frac{1}{c^2}\frac{\partial^2\psi}{\partial t^2}-\Delta\psi = 0,\qquad \psi\in\Omega\times\mathbb{R},\qquad\Omega\subset\mathbb{R}^2,
  \label{eq:2dwave}
\end{equation}
where $\Delta=\partial_x^2+\partial_y^2$, $c$ is the speed of propagation, and $\Omega$ is a bounded domain in $\mathbb{R}^2$.
The product ansatz $\psi(t,x,y)=v(t)\cdot u(x,y)$ immediately yields the two differential equations
\begin{equation}
  \partial_t^2v + \lambda c^2v = 0\quad\mbox{and}\quad \Delta u + \lambda u = 0
  \label{eq:2dwavesep}
\end{equation}
with a positive constant $\lambda$. The solution to the ordinary differential equation for $v(t)$ in \eref{eq:2dwavesep} has the form $v(t)=a\cos(\omega t)+b\sin(\omega t)$ with $\omega=c\sqrt{\lambda}$. The partial differential equation for $u(x,y)$ in \eref{eq:2dwavesep} leads to the eigenvalue problem of the Laplace operator we have to solve.

Now, the idea of the finite element approach is to use the equivalent variational integral formulation of the eigenvalue system and to decompose the integration domain $\Omega$ into finite elements like, for example, a triangular mesh. The variational integral for the eigenvalue system reads
\begin{equation}
  \mathcal{I} = \frac{1}{2}\iint_{\Omega}\left\{\left[\left(\partial_xu\right)^2+\left(\partial_yu\right)^2\right]-\lambda u^2\right\}dx\;dy+\frac{1}{2}\int\limits_{\partial\Omega}u^2ds,
  \label{eq:varInt}
\end{equation}
where $\partial\Omega$ are all boundary curve segments that are supported elastically. If the boundary of $\Omega$ is fixed or free, then this integral can be dropped.
The domain $\Omega$ is approximated by the union of all triangle elements $T_i = \triangle P_{i_1}P_{i_2}P_{i_3}$, $\bigcup_{i=1}^N T_i\subseteq\Omega$. The indices $(i_1,i_2,i_3)$ represent the numbers of the mesh vertices that make up the $i$-th triangle.

\begin{figure}[ht]
  \includegraphics[scale=0.8]{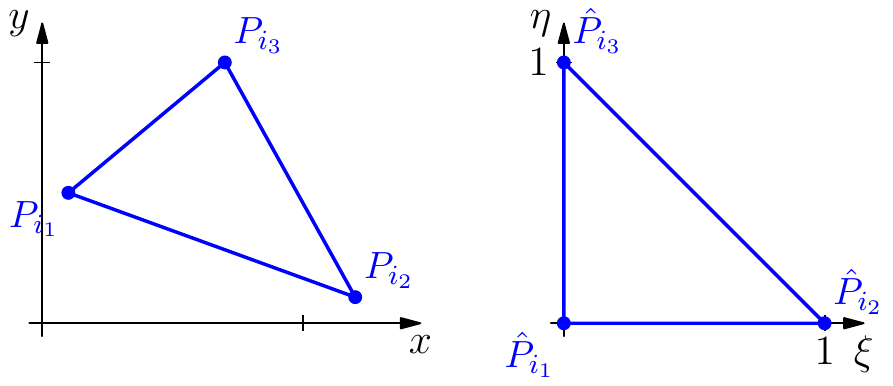}
  \caption{Transformation of a triangle $\triangle P_{i_1}P_{i_2}P_{i_3}$ from Cartesian coordinates $(x,y)$ to the standard triangle $\triangle\hat{P}_{i_1}\hat{P}_{i_2}\hat{P}_{i_3}$ in canonical coordinates $(\xi,\eta)$.}
  \label{fig:triTransf}
\end{figure}

Within each triangle element, the function $u$ is approximated by a linear or a higher order ansatz function $u_i$ that vanishes outside $T_i$. Hence, $u(x,y)\approx\sum_{i=1}^Nu_i(x,y)$, and the integral in \eref{eq:varInt} separates into integrals $\mathcal{I}_i$ over the single elements. To normalize these integrals, they are transformed to canonical coordinates via
\begin{eqnarray}
    x &= x_{i_1} + (x_{i_2}-x_{i_1})\xi + (x_{i_3}-x_{i_1})\eta,\\
    y &= y_{i_1} + (y_{i_2}-y_{i_1})\xi + (y_{i_3}-y_{i_1})\eta,
\end{eqnarray}
see also Fig.~\ref{fig:triTransf}.

The derivatives of the ansatz functions $u_i$ transform as follows
\begin{eqnarray}
    \partial_xu_i &= (\partial_{\xi}u_i)(\partial_x\xi) + (\partial_{\eta}u_i)(\partial_x\eta) = \frac{y_{i_3}-y_{i_1}}{J_i}\partial_{\xi}u_i - \frac{y_{i_2}-y_{i_1}}{J_i}\partial_{\eta}u_i,\\
    \partial_yu_i &= (\partial_{\xi}u_i)(\partial_y\xi) + (\partial_{\eta}u_i)(\partial_y\eta) = -\frac{x_{i_3}-x_{i_1}}{J_i}\partial_{\xi}u_i + \frac{x_{i_2}-x_{i_1}}{J_i}\partial_{\eta}u_i
\end{eqnarray}
with the determinant of the Jacobian,  $J_i=\det\left[\partial(x,y)/\partial(\xi,\eta)\right]=(x_{i_2}-x_{i_1})(y_{i_3}-y_{i_1})-(x_{i_3}-x_{i_1})(y_{i_2}-y_{i_1})$.
Thus, the first part of the variational integral over the triangle $T_i$ transforms to
\begin{eqnarray}
   \nonumber\fl 2\mathcal{I}_i=\iint_{T_i}\left\{\left[(\partial_xu_i)^2+(\partial_yu_i)^2\right]-\lambda u_i^2\right\}\mathrm{d}x \mathrm{d}y \\
   \label{eq:varInts}\fl = a_i\iint_{T_i}(\partial_{\xi}u_i)^2 d\xi d\eta + b_i\iint_{T_i}2(\partial_{\xi}u_i)(\partial_{\eta}u_i)d\xi d\eta + c_i\iint_{T_i}(\partial_{\eta}u_i)^2 d\xi d\eta - \lambda J_i\iint_{T_i}u_i^2d\xi d\eta,\\
   \label{eq:varIntTrnsf}\fl = a_i I_{(1)i} + b_i I_{(2)i} + c_i I_{(3)i} - \lambda J_i I_{(4)i}.
\end{eqnarray}
with triangle depending constants
\begin{eqnarray}
  a_i &= \left[(x_{i_3}-x_{i_1})^2 + (y_{i_3}-y_{i_1})^2\right]/J_i,\\
  b_i &= -\left[(x_{i_3}-x_{i_1})(x_{i_2}-x_{i_1}) + (y_{i_3}-y_{i_1})(y_{i_2}-y_{i_1})\right]/J_i,\\
  c_i &= \left[(x_{i_2}-x_{i_1})^2 + (y_{i_2}-y_{i_1})^2\right]/J_i.
\end{eqnarray}

In case of the linear ansatz for the function $u_i$, where $u_i(x,y)=\upsilon_{(1)i}+\upsilon_{(2)i}x+\upsilon_{(3)i}y$ or $u_i(\xi,\eta)=\alpha_{(1)i}+\alpha_{(2)i}\xi+\alpha_{(3)i}\eta$, respectively, the integrals in \eref{eq:varInts} can be easily evaluated, and the three values $(u_{i_1},u_{i_2},u_{i_3})=(\alpha_{(1)i},\alpha_{(1)i}+\alpha_{(2)i},\alpha_{(1)i}+\alpha_{(3)i})$ at the corners $P_{i_1}$, $P_{i_2}$, and $P_{i_3}$ completely determine the integrals $I_{(n)i}$
In matrix notation, these integrals can be written as $I_{(n)i}=\vec{\alpha}_i^T\tilde{S}_{(n)}\vec{\alpha}_i$, where $\vec{\alpha}_i^T=(\alpha_{(1)i},\alpha_{(2)i},\alpha_{(3)i})^T$ and
\begin{eqnarray}
  \tilde{S}_{(1)} &= \frac{1}{2}\left(\begin{array}{ccc}0&0&0\\0&1&0\\0&0&0\end{array}\right),\quad \tilde{S}_{(2)} = \frac{1}{2}\left(\begin{array}{ccc}0&0&0\\0&0&1\\0&1&0\end{array}\right),\\ 
  \tilde{S}_{(3)} &= \frac{1}{2}\left(\begin{array}{ccc}0&0&0\\0&0&0\\0&0&1\end{array}\right),\quad \tilde{S}_{(4)} = \frac{1}{24}\left(\begin{array}{ccc}12&4&4\\4&2&1\\4&1&2\end{array}\right).
\end{eqnarray}
For further calculations, the integrals have to be related to the values at the corners,   $\vec{u}_i=(u_{i_1},u_{i_2},u_{i_3})^T$. Then,  $I_{(n)i}=\vec{u}_i^TA^T\tilde{S}_{(n)}A\vec{u}_i=\vec{u}_i^TS_{(n)}\vec{u}_i$ with
\begin{eqnarray}
  \fl S_{(1)} &= \frac{1}{2}\left(\begin{array}{rrr}1&-1&0\\-1&1&0\\0&0&0\end{array}\right),\quad S_{(2)} = \frac{1}{2}\left(\begin{array}{rrr}2&-1&-1\\-1&0&1\\-1&1&0\end{array}\right),\\
  \fl S_{(3)} &= \frac{1}{2}\left(\begin{array}{rrr}1&0&-1\\0&0&0\\-1&0&1\end{array}\right),\quad
  S_{(4)} = \frac{1}{2}\left(\begin{array}{rrr}2&1&1\\1&2&1\\1&1&2\end{array}\right),\quad
  A = \left(\begin{array}{rrr}1&0&0\\-1&1&0\\-1&0&1\end{array}\right).
\end{eqnarray}
Finally, the variational integral over the triangle $T_i$ is given by $2\mathcal{I}_i=\vec{u}_i^T\left(S_i-\lambda M_i\right)\vec{u}_i$, where $S_i=a_iS_{(1)}+b_iS_{(2)}+c_iS_{(3)}$ is called the \emph{stiffness element matrix} for the element $i$, and $M_i=J_iS_{(4)}$ is its \emph{mass element matrix}.

In case of elastically supported edges, the boundary segment integral in equation \eref{eq:varInt} has to be taken into account. With the linear ansatz $u_{i}=u_{i_1}(1-\xi)+u_{i_2}\xi$ for the edge $(i_1,i_2)$ of triangle $i$, this integral reads
\begin{equation}
    I_{(5)i}=\int_{T_i}u_i^2ds = l_{i_1i_2}\int_{T_i}\left[u_{i_1}(1-\xi)+u_{i_2}\xi\right]d\xi,\qquad l_{i_1i_2}=\overline{P_{i_1}P_{i_2}},
\end{equation}
where $l_{i_1i_2}$ is the length of the edge between $P_{i_1}$ and $P_{i_2}$.
In matrix notation, $I_{(5)i}$ can also be written as
\begin{eqnarray}
  I_{(5)i}=l_{i_1i_2}\left(\begin{array}{rr}u_{i_1}&u_{i_2}\end{array}\right)S_{(5)}\left(\begin{array}{r}u_{i_1}\\ u_{i_2}\end{array}\right)\quad\mbox{with}\quad S_{(5)} = \frac{1}{6}\left(\begin{array}{rr}2&1\\1&2\end{array}\right),
  \label{eq:boundSegInt}
\end{eqnarray}
which is also valid for any other of the triangle edges.

Now, the complete system is obtained by compiling the stiffness matrix $\mathcal{S}$ and the mass matrix $\mathcal{M}$ from the element matrices, where both matrices are of size $N\times N$. The unknown function values $u_{i_n}$ at the vertices of the triangular mesh build the $N$-dimensional vector $\underline{u} = \left(u_{1},u_{2},\ldots,u_{N}\right)^T$. Then, for each triangle $i$ with vertex indices $(i_1,i_2,i_3)$ and $i_1,i_2,i_3\in[1,N]$, the matrix entries have to be summed up,
\begin{equation}
   (\mathcal{S})_{i_ni_m} \mathrel{+}= (S_i)_{nm},\qquad (\mathcal{M})_{i_ni_m} \mathrel{+}= (M_i)_{nm},\qquad n,m\in\{1,2,3\}.
   \label{eq:compile}
\end{equation}
Equation \eref{eq:compile} is to be read as follows: first, the matrices $\mathcal{S}$ and $\mathcal{M}$ must be initialized to zero. Next, for each triangle $i$, the associated vertex indices $(i_1,i_2,i_3)$ have to be determined. At the corresponding matrix positions $(i_ni_m)$, the component $(nm)$ of the element matrix determined for that triangle must be added. The elastically supported boundary segments, Eq.~\eref{eq:boundSegInt}, have to be added to $\mathcal{S}$ in the similar way.

After all element matrices were added, those rows and columns for which the mesh vertices are fixed can be deleted because these values are already known. The final step is the variation of the integral \eref{eq:varInt}, $\delta\mathcal{I}=0$, which yields the eigenvalue system
\begin{equation}
  \mathcal{S}\underline{u} = \lambda \mathcal{M}\underline{u}
  \label{eq:evs}
\end{equation}
that has to be solved. For every resulting eigenvalue $\lambda$, the eigenvector $\underline{u}$ determines the maximum displacements of the non-fixed triangle mesh vertices.

%
\section{The graphical user interface \emph{NumChladni}}\label{sec:numchladni}
In contrast to experimental studies of Chladni figures with prefabricated metal plates, the computational simulation offers the possibility to explore in detail arbitrarily shaped membranes.
\emph{NumChladni} is a graphical user interface that makes this exploration interactive and comfortable.

\begin{figure}[ht]
   \includegraphics[scale=0.28]{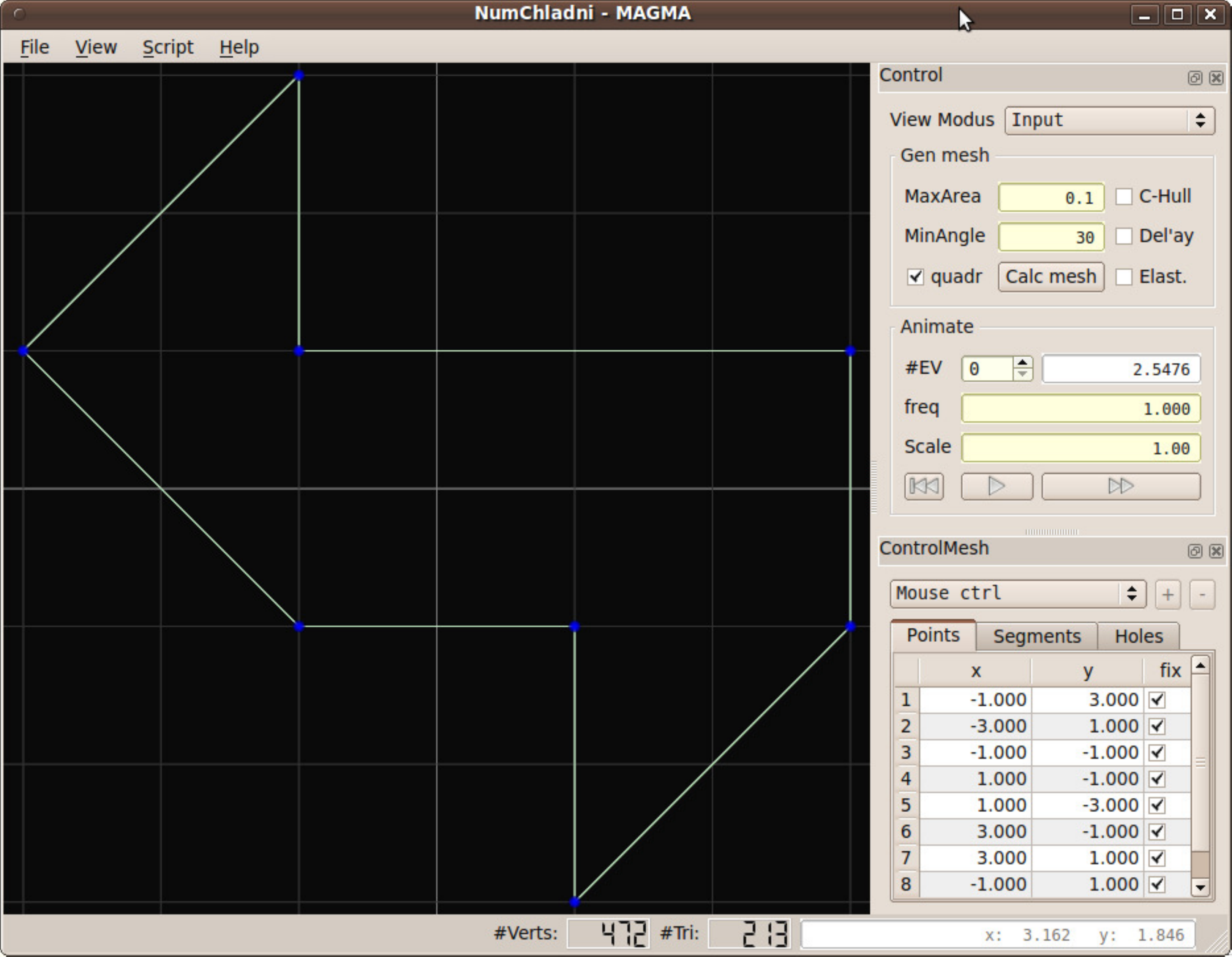}\,\,\includegraphics[scale=0.3]{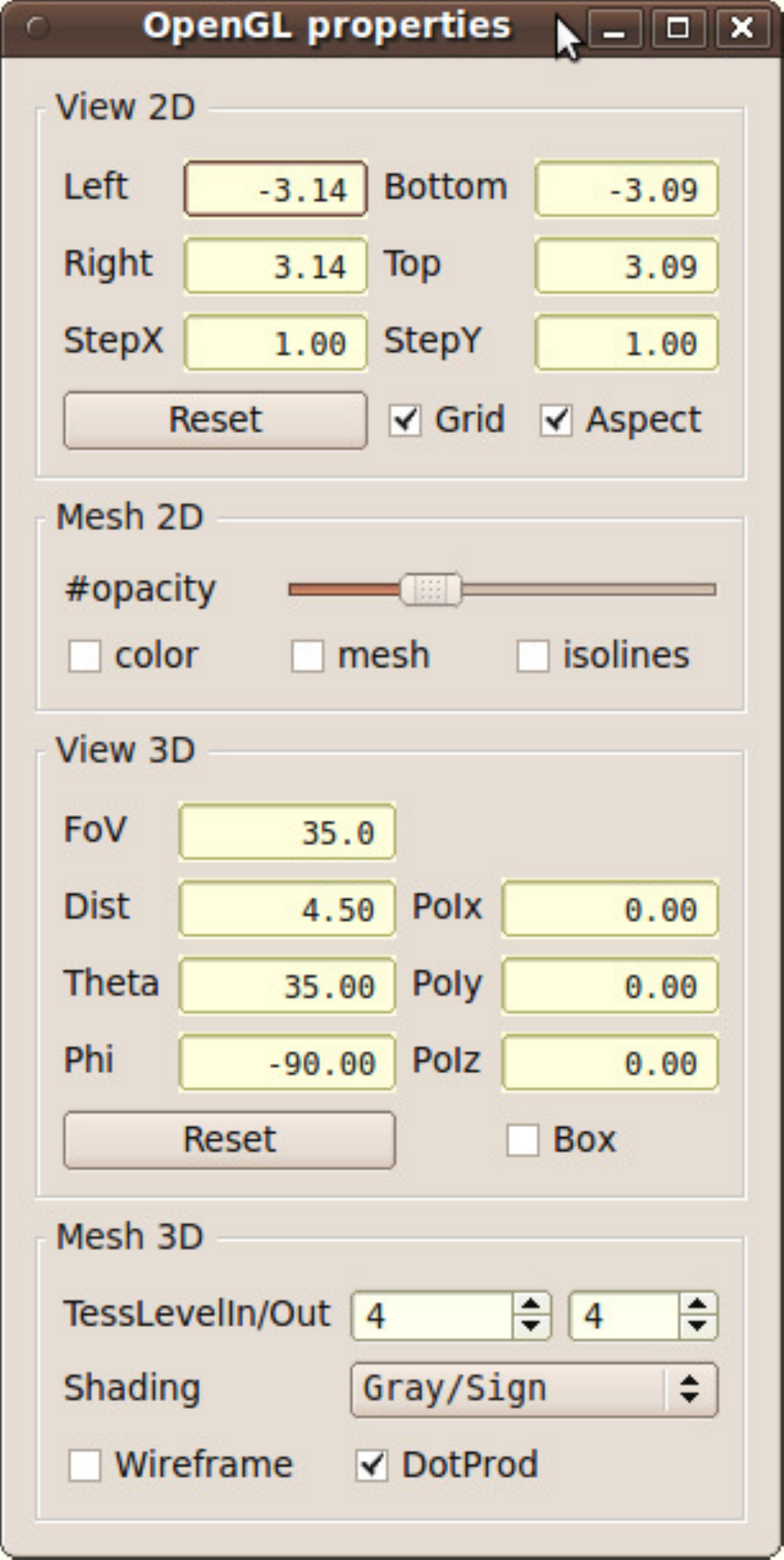}
   \caption{Screenshots of \emph{NumChladni}'s graphical user interface which is based on the Qt framework~\cite{QT} and the Open Graphics Library (OpenGL)\cite{opengl}. In the ControlMesh tab, points and segments can also be entered in table view. The enclosed region represents the polygonal shape $\Omega$. Holes can be realized by setting a point inside a sub polygon. In 2D and 3D view modus, the eigenmodes to the eigenvalues (\#EV) can be animated.}
   \label{fig:ncScreenshot}
\end{figure}

\emph{NumChladni} starts in the input view mode with an empty drawing board. To enter a new polygonal mesh, you first have to add some points which can be joined by segments afterwards. For every point, you can decide whether it is fixed or it can move freely. An edge is fixed, if both ends are fixed. All edges that are not fixed are either supported elastically or are free. Figure~\ref{fig:ncScreenshot} shows a screenshot of \emph{NumChladni} after this first step. The polygonal mesh can also be stored for later use.

When the polygonal shape is set, a triangular mesh must be generated. This meshing is a non-trivial task and a detailed discussion is out of the scope of this article. In \emph{NumChladni}, it is delegated to Shewchuk's triangle library~\cite{triangle}. The quality of the mesh can be influenced by several parameters: the maximum area of a triangle element, the minimum angle within the triangle, whether the mesh should be Delaunay triangulated or not, and the degree of the ansatz function being either linear or quadratic.
After the meshing step, the stiffness and mass matrices are compiled, and the eigenvalue system \eref{eq:evs} is solved using either the Gnu Scientific Library (GSL)~\cite{gsl}, the LAPACK~\cite{lapack} package, or the MAGMA~\cite{magma} library that makes use of the large compute capability of graphic boards.
\begin{figure}[ht]
  \includegraphics[width=0.45\textwidth]{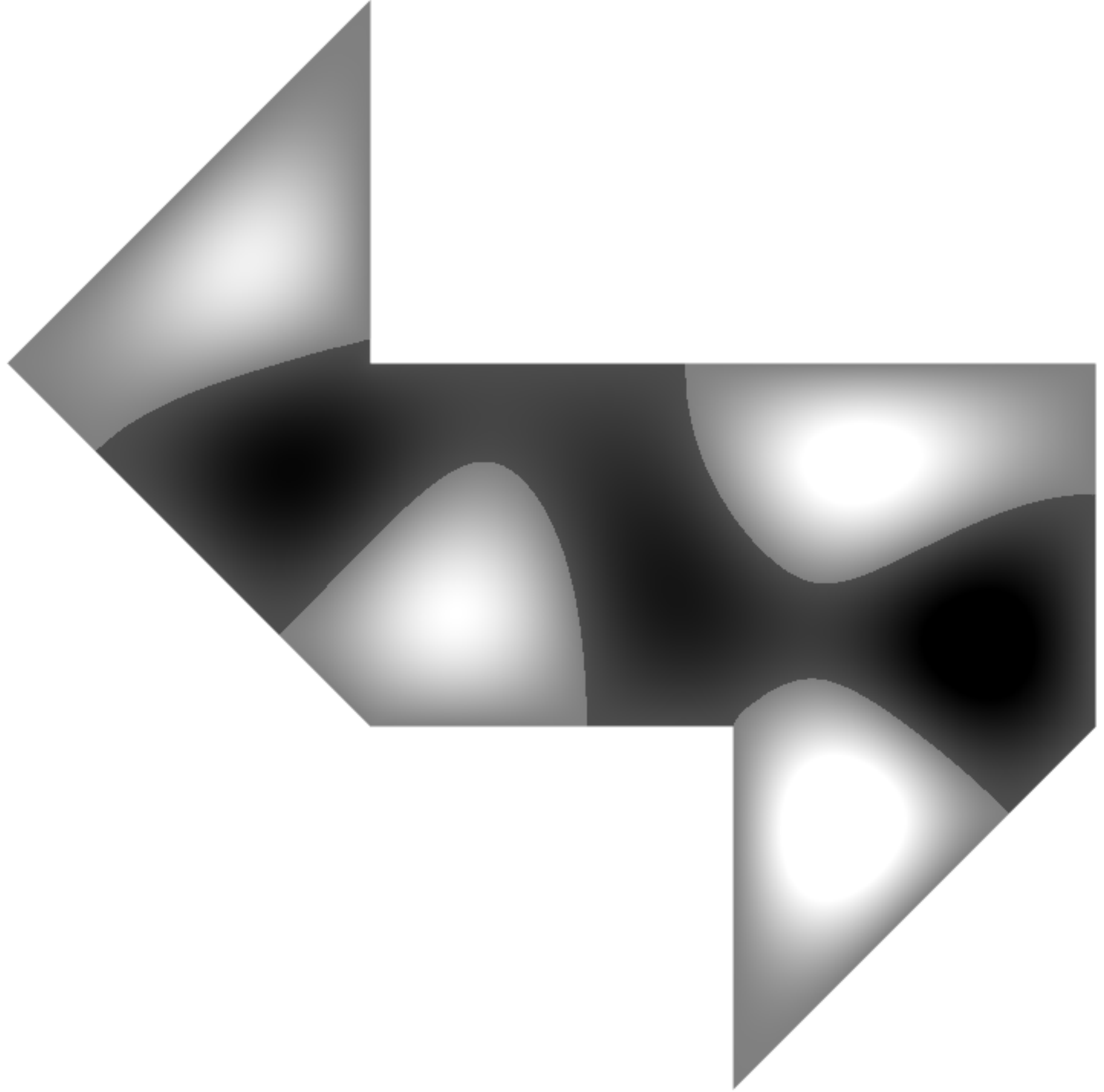}\hspace*{1em}
  \includegraphics[width=0.53\textwidth]{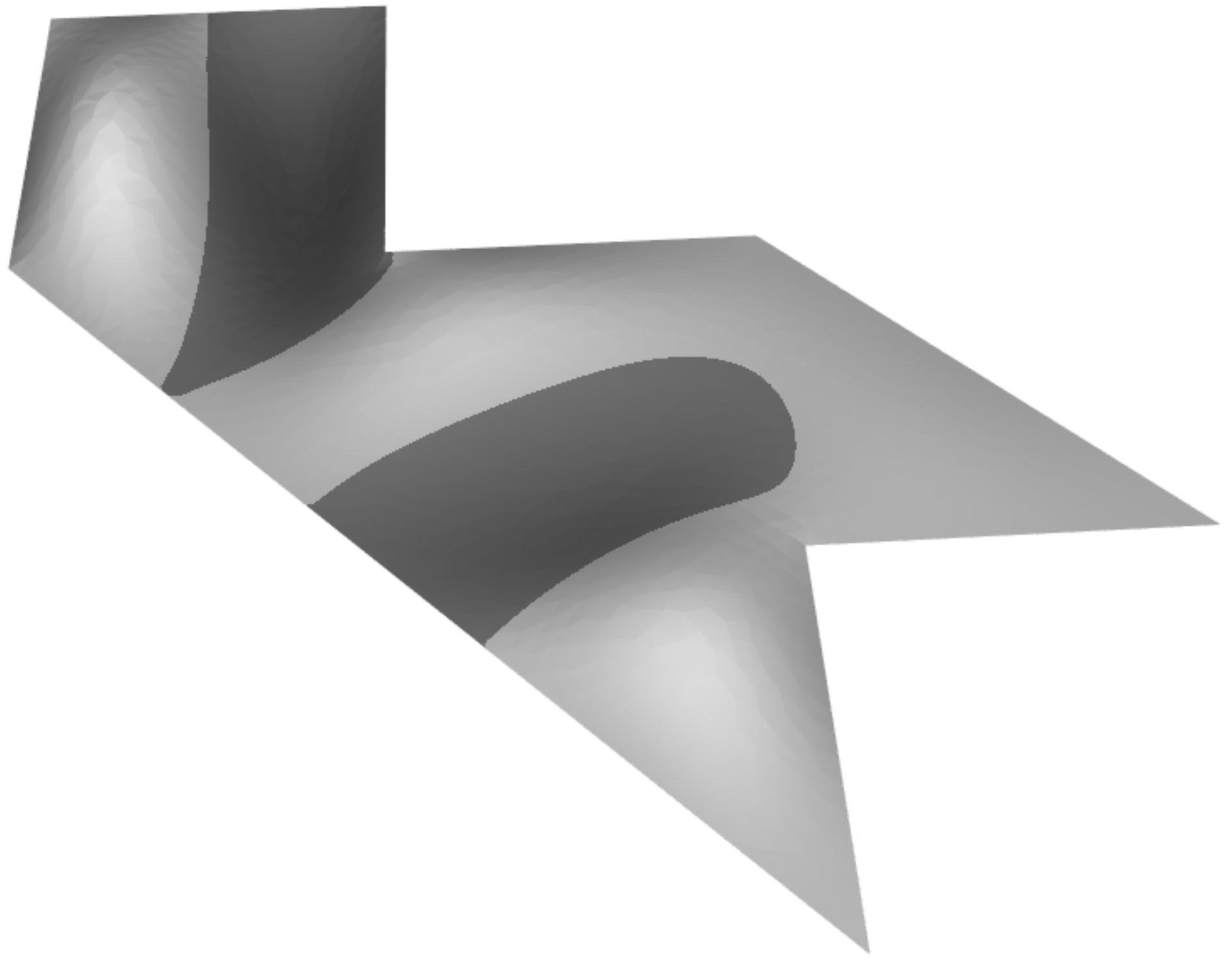}
  \caption{Screenshots of \emph{NumChladni}'s 2D (left) and 3D (right) projection. Here, two isospectral drums with $\lambda_7=11.54$ are shown, see Driscoll~\cite{Driscoll1997} for details. The nodal lines are represented by the transition lines between light and dark gray regions.}
  \label{fig:ncDrum}
\end{figure}

The resulting eigenmodes can be displayed in 2D and 3D projection, see Fig.~\ref{fig:ncDrum}. In both view modi, the eigenmodes can be animated by a simple cosine function where the eigen frequency $\omega=c\sqrt{\lambda}$ is neglected but is set to unity. The nodal lines are indicated by the transition lines between light and dark gray regions. On high-end graphic boards, the triangle mesh can be subdivided using tessellation shaders to obtain a smoother surface.

To simplify the input of some more complex shapes like, for example, $N$-gones, and for better reproducability, \emph{NumChladni} is scriptable. Hence, entering a polygon, setting the mesh parameters, and adjusting the view parameters can be merged into one script. Owing to the mathematical functions that are available in the scripting environment, even circular discs or ellipses could be easily generated.

%
\section{Examples}\label{sec:examples}
\begin{figure}[ht]
  \includegraphics[width=0.24\textwidth]{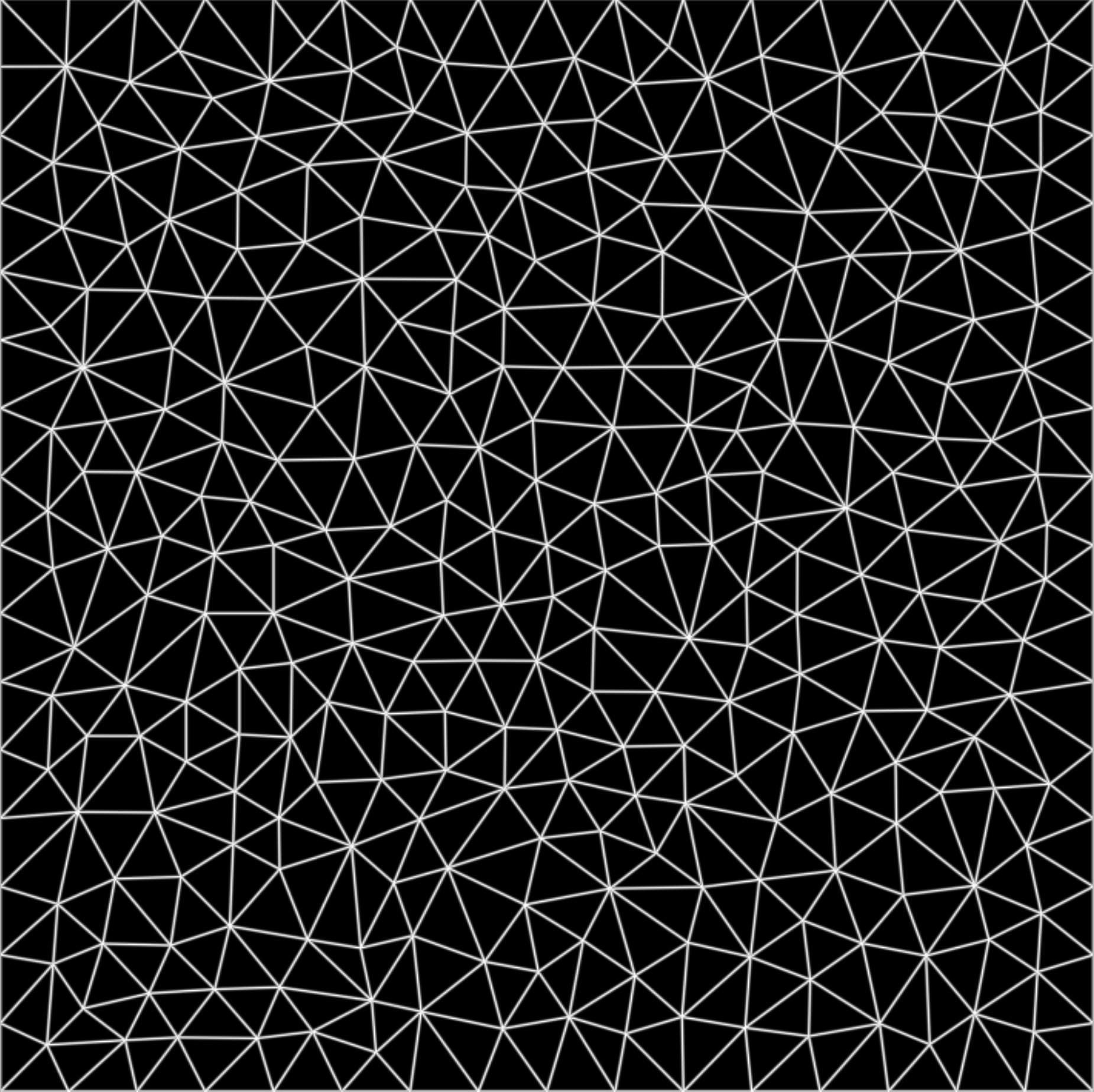}\,%
  \includegraphics[width=0.24\textwidth]{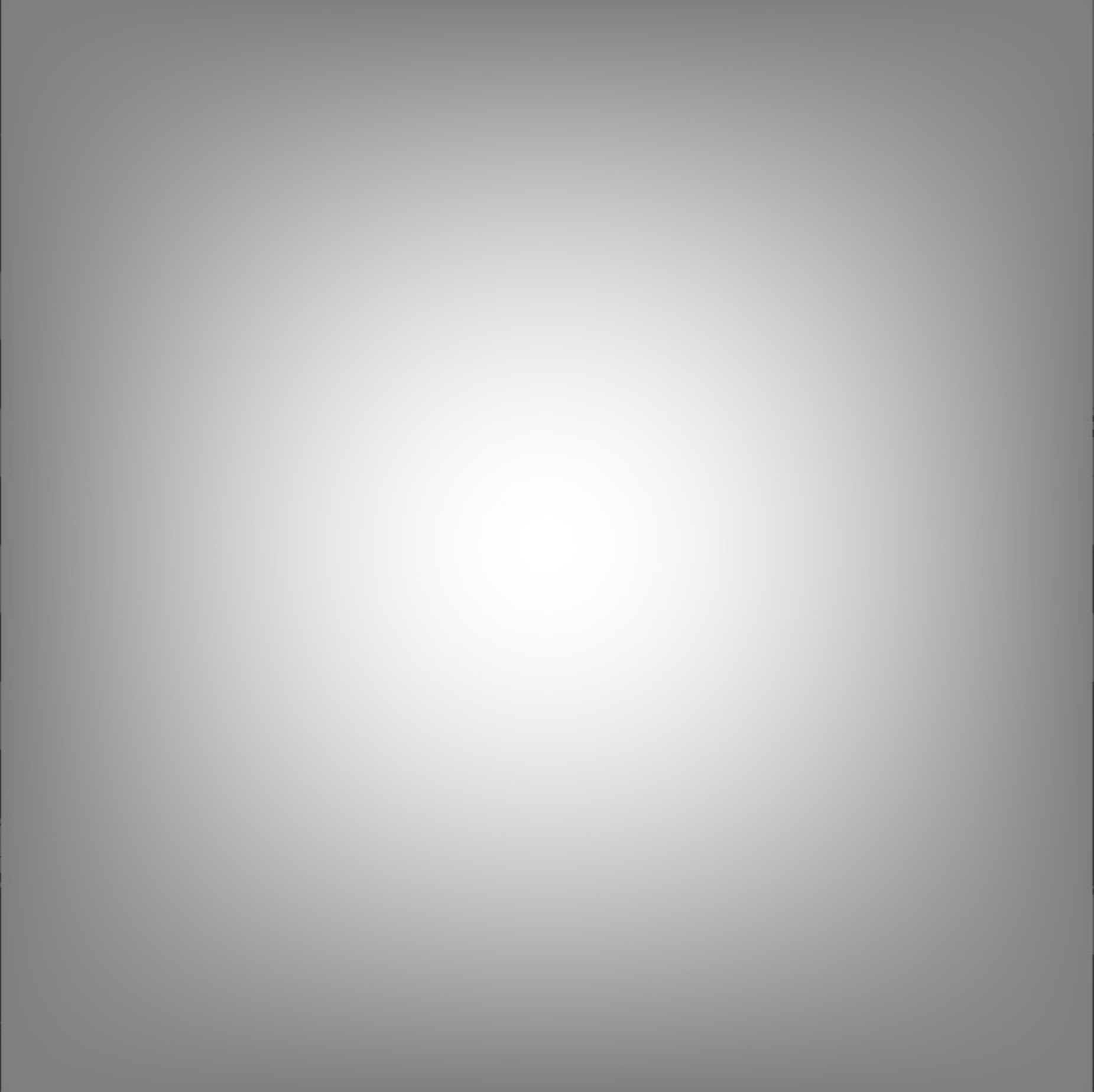}\,%
  \includegraphics[width=0.24\textwidth]{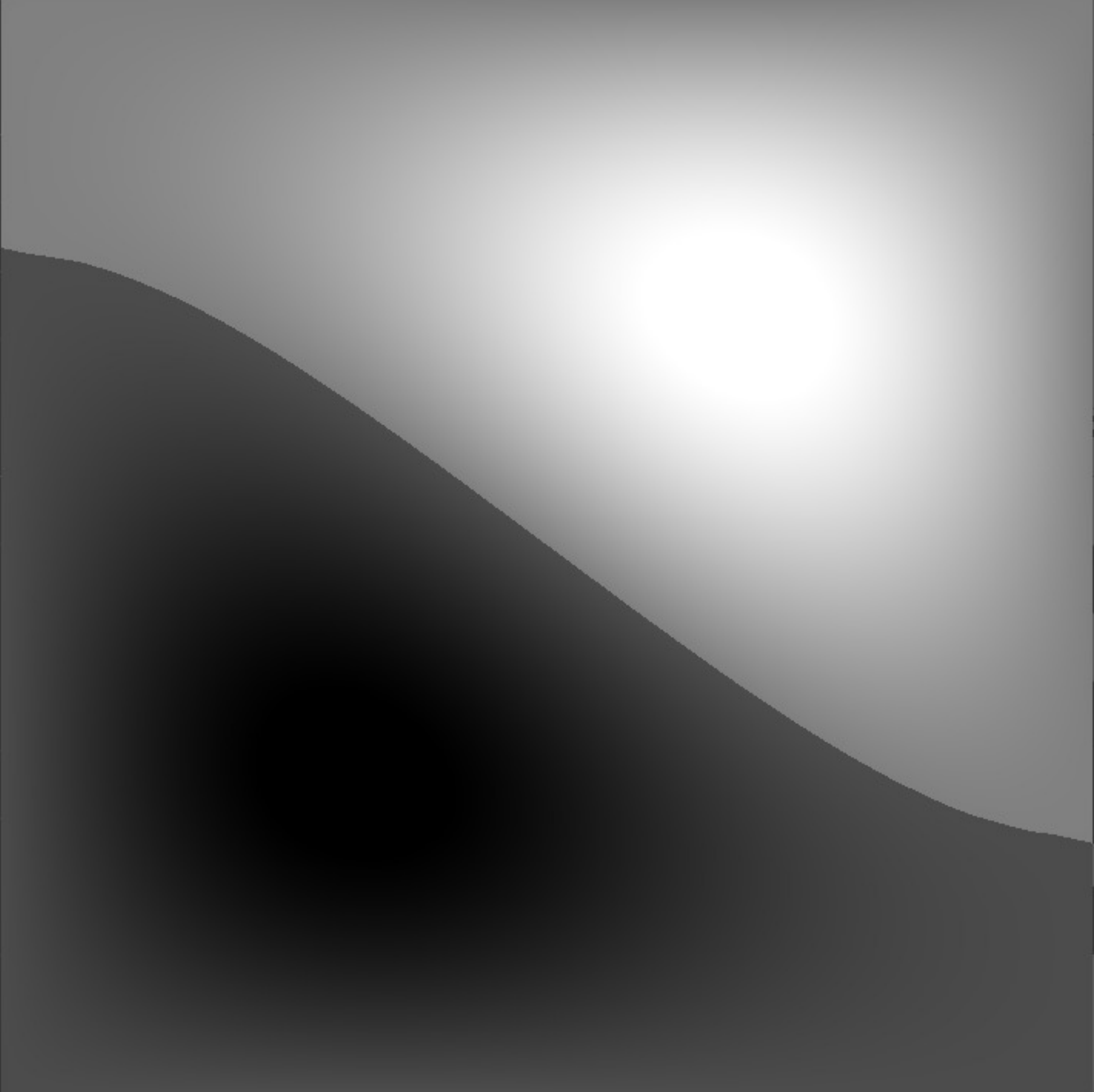}\,%
  \includegraphics[width=0.24\textwidth]{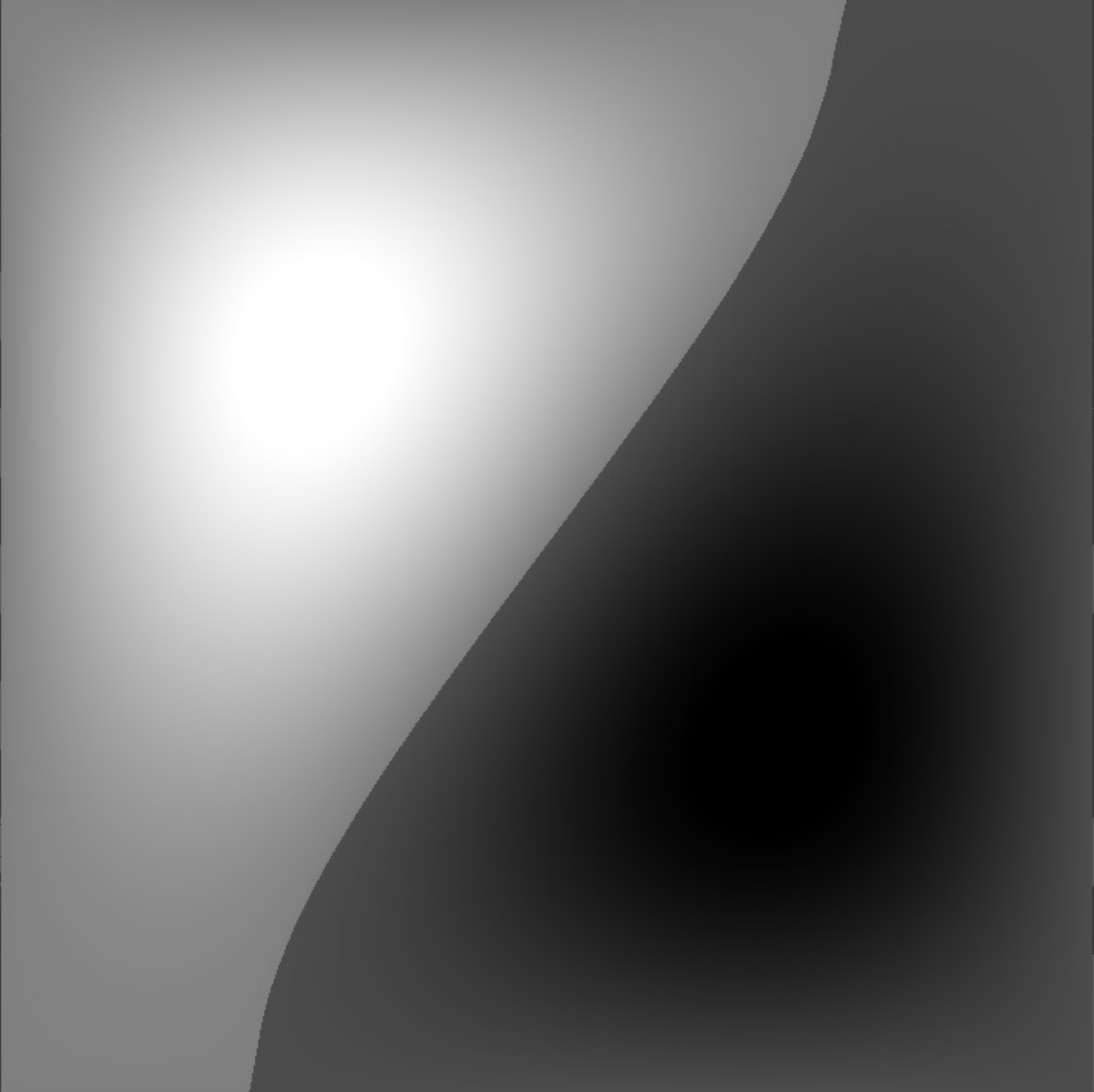}\\
  \includegraphics[width=0.24\textwidth]{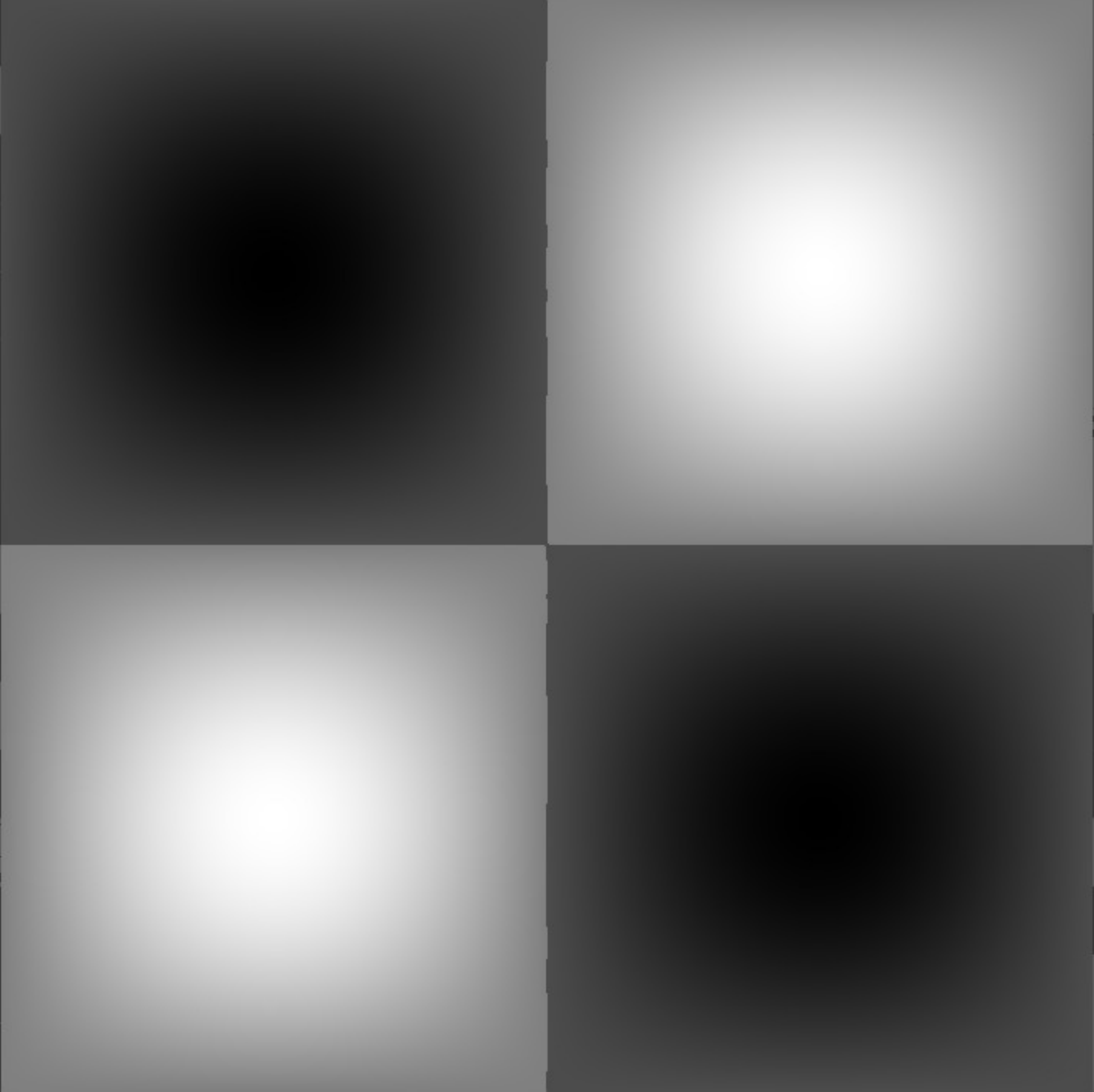}\,%
  \includegraphics[width=0.24\textwidth]{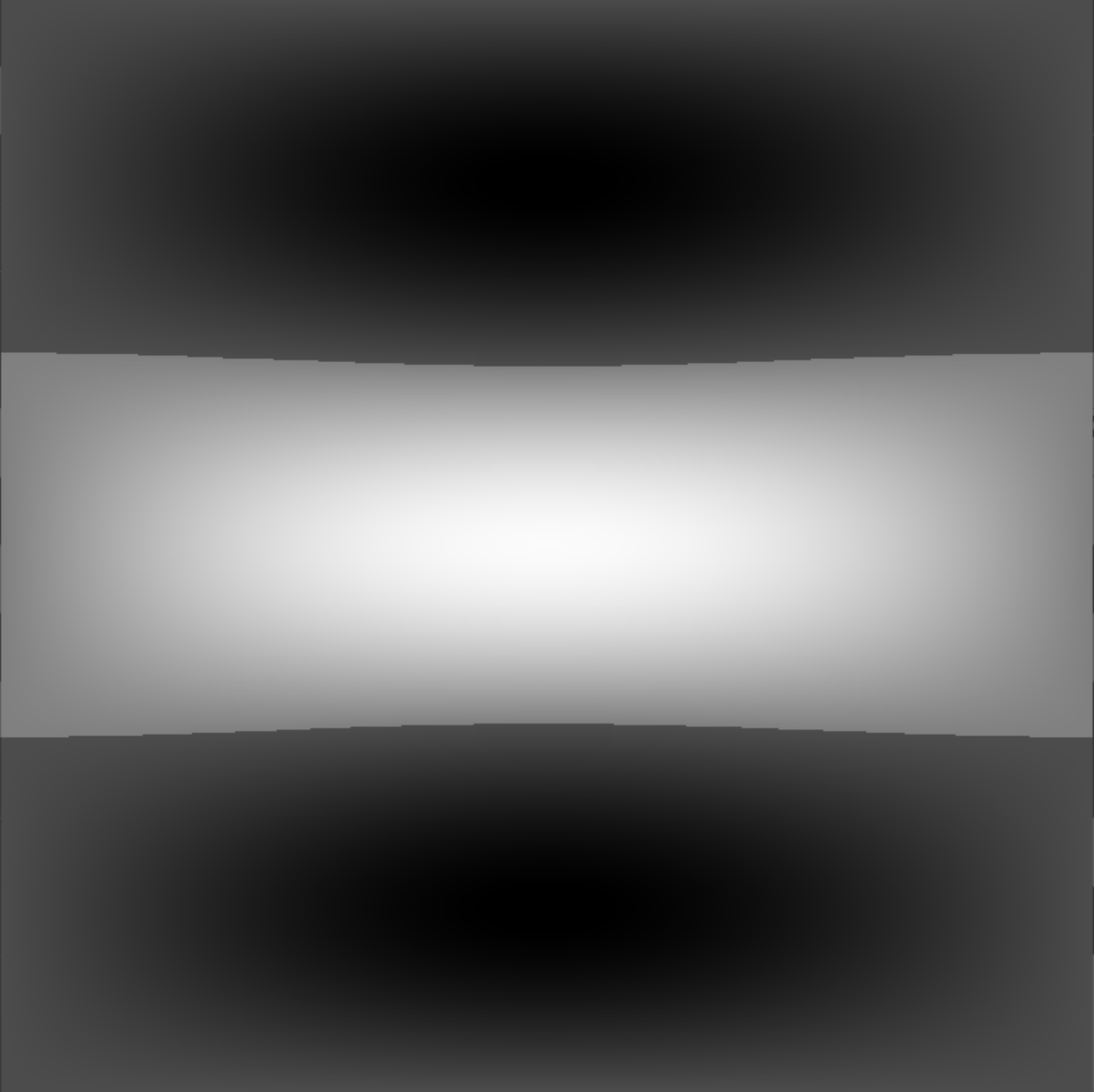}\,%
  \includegraphics[width=0.24\textwidth]{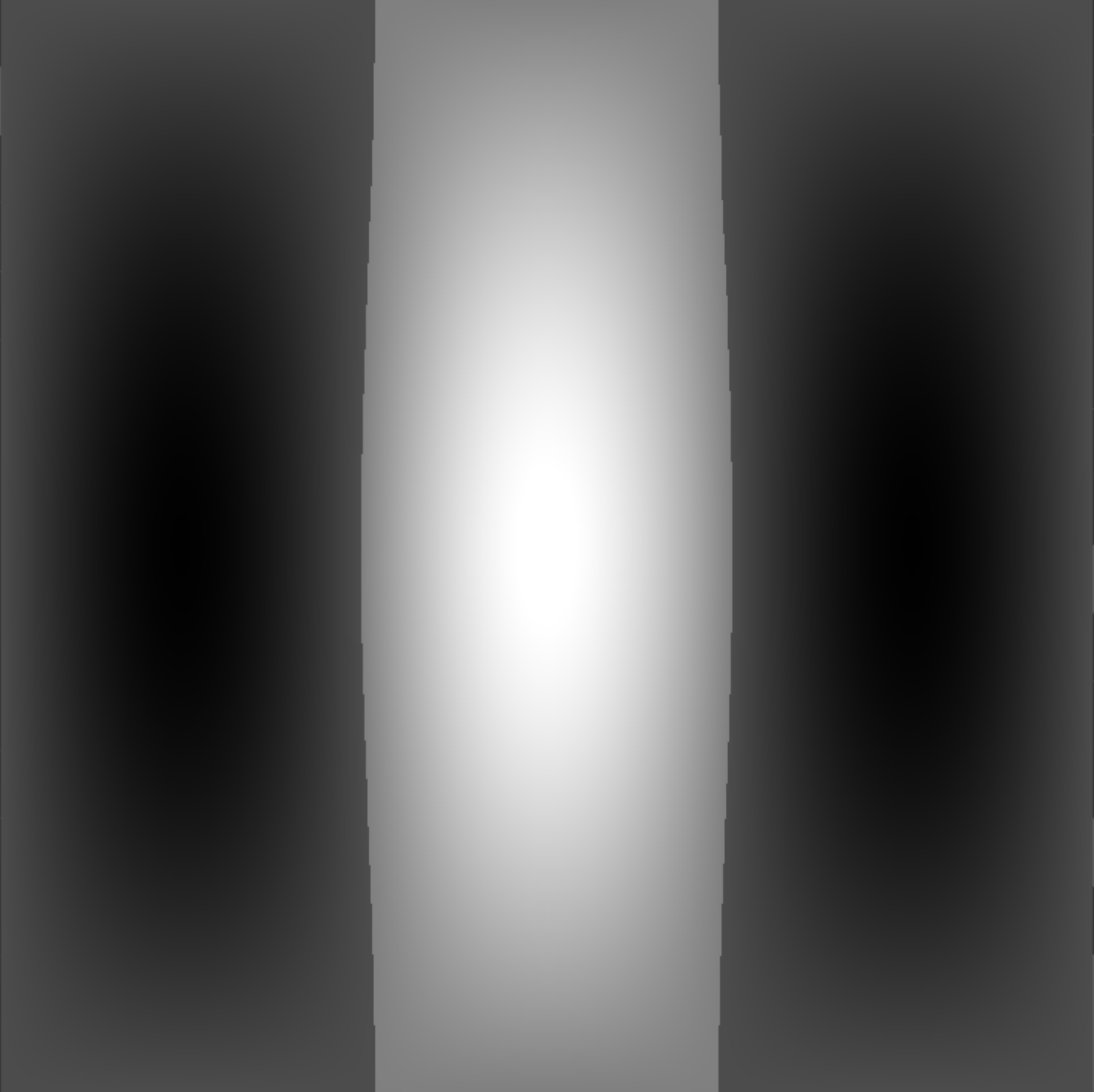}\,%
  \includegraphics[width=0.24\textwidth]{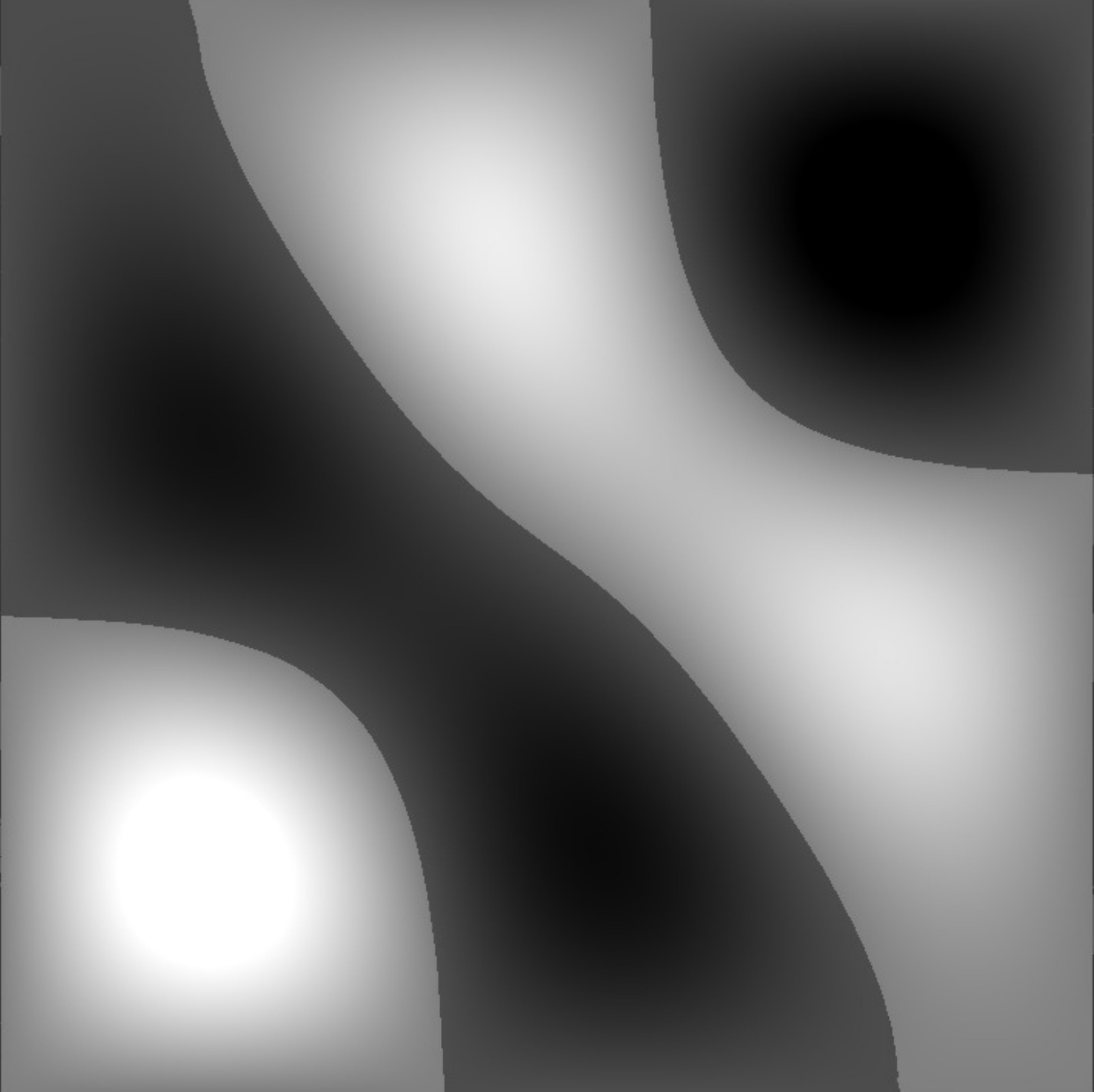}\\
  \includegraphics[width=0.24\textwidth]{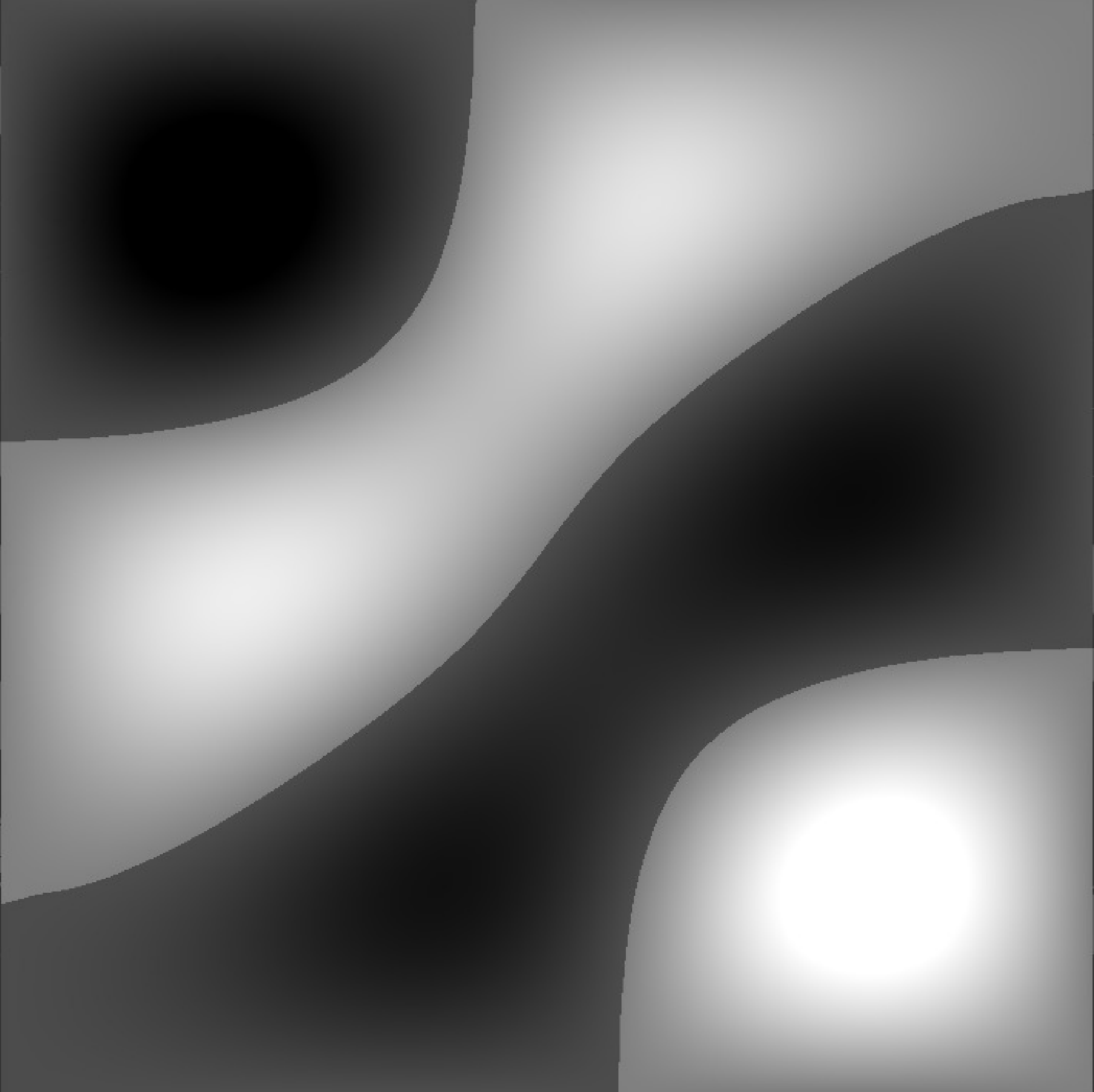}\,%
  \includegraphics[width=0.24\textwidth]{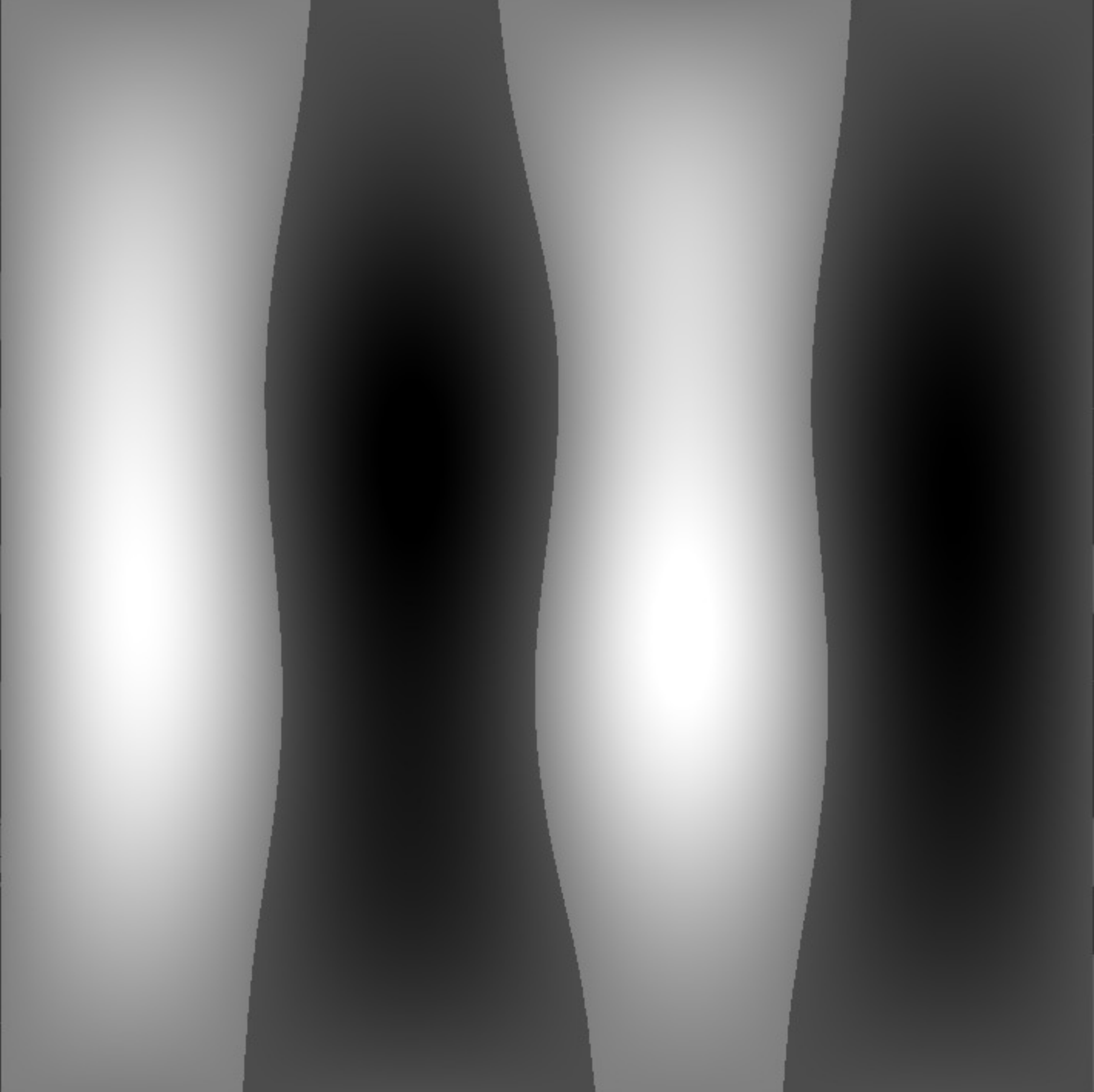}\,%
  \includegraphics[width=0.24\textwidth]{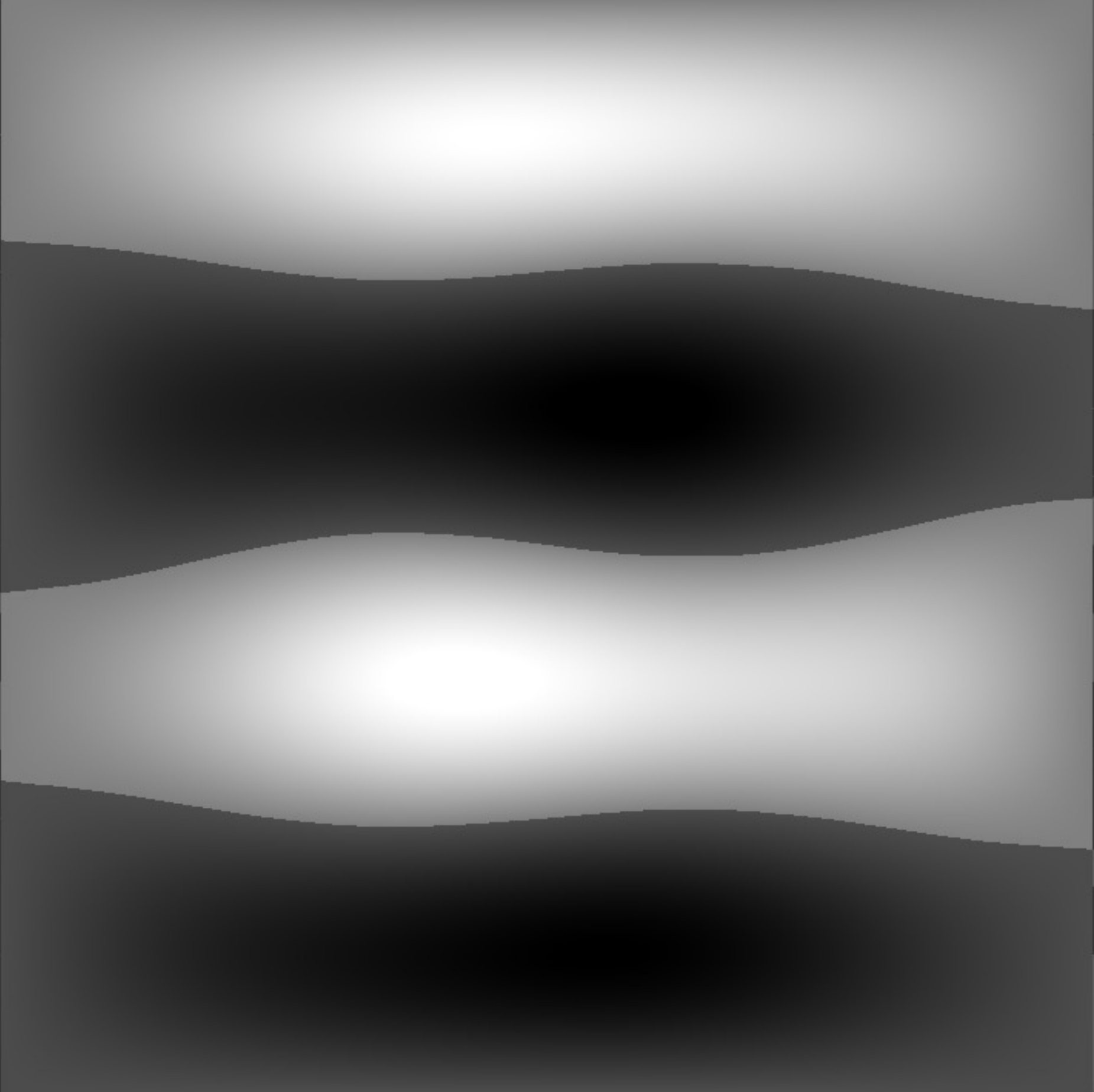}\,%
  \includegraphics[width=0.24\textwidth]{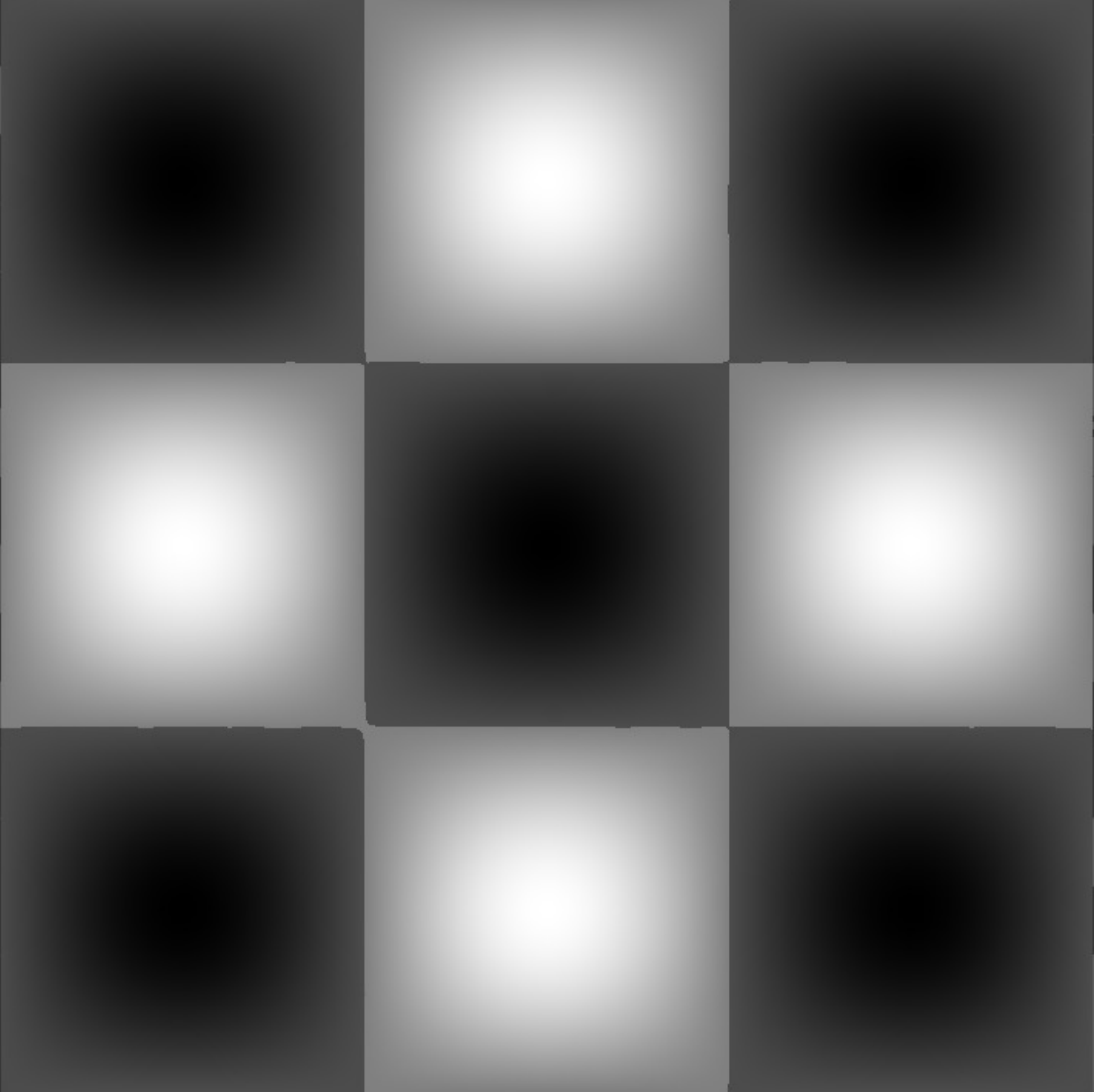}
  \caption{The mesh for the quadratic membrane (upper left image, edge length $L=2$) consists of $1313$ vertices making up $624$ triangles with quadratic ansatz. It was generated with parameters $maxArea=0.01$ and $minAngle=30$. The boundary is fixed. The eigenvalues read: $\lambda=\{4.9348, 12.3375, 12.3375, 19.7413, 24.6779, 24,6783, 32.0850, 32.0853, 41.9637,$ $41.9642, 49.3786\}$ (upper left to lower right).}
  \label{fig:quadImages}
\end{figure}

The most simple example that could be solved also analytically in a straightforward manner is the rectangular membrane where the boundary is fixed, see \ref{appsec:quadMembrane}. The eigenmodes of the special case of a quadratic membrane with edge lengths $a=b=2$ determined with \emph{NumChladni} are shown in Fig.~\ref{fig:quadImages}. However, because of the symmetric situation, the eigenmodes are degenerated and the FE method does not show the `natural' modes but some mixed frequency solutions.

\begin{figure}[ht]
  \includegraphics[width=0.24\textwidth]{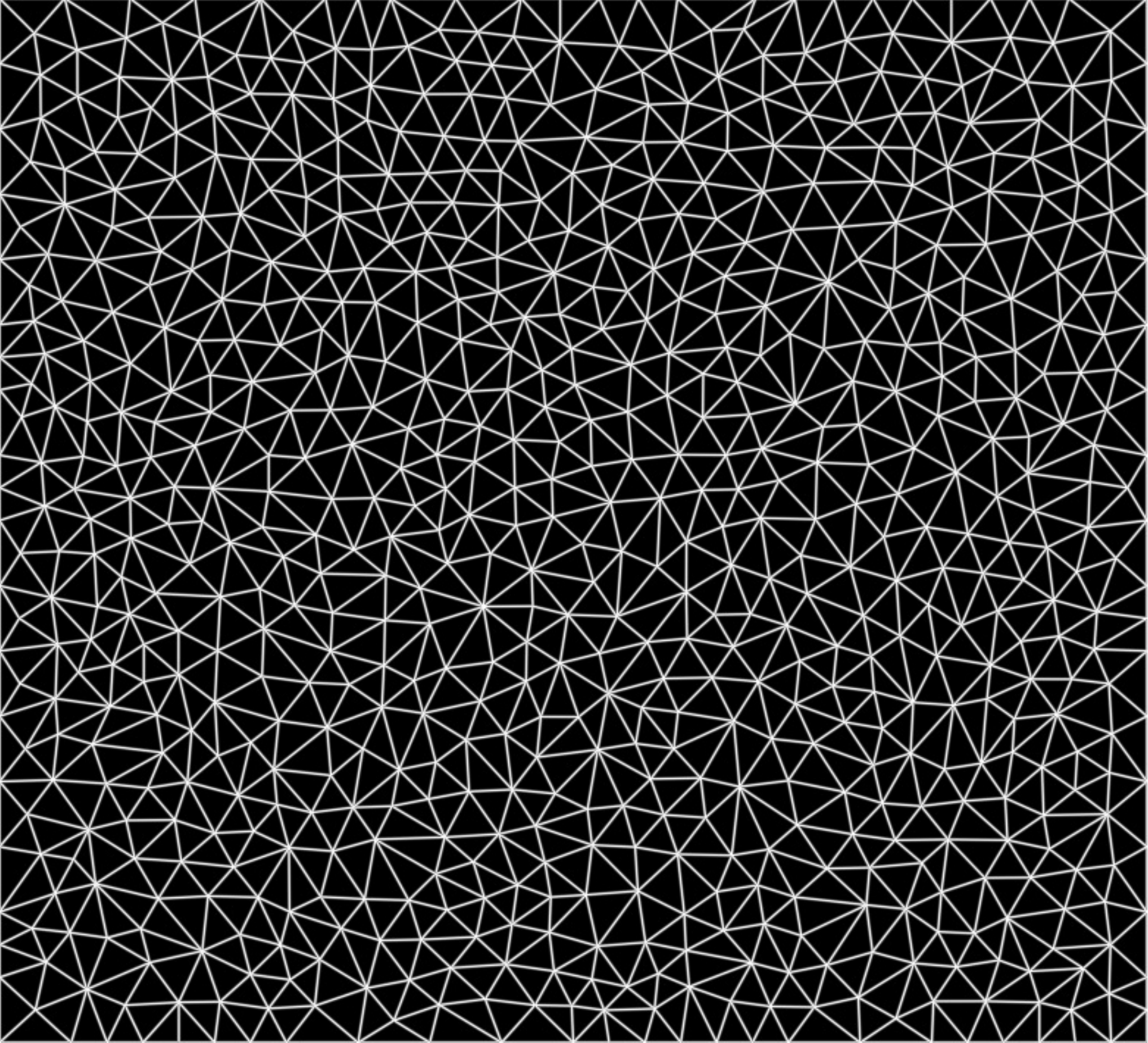}\,%
  \includegraphics[width=0.24\textwidth]{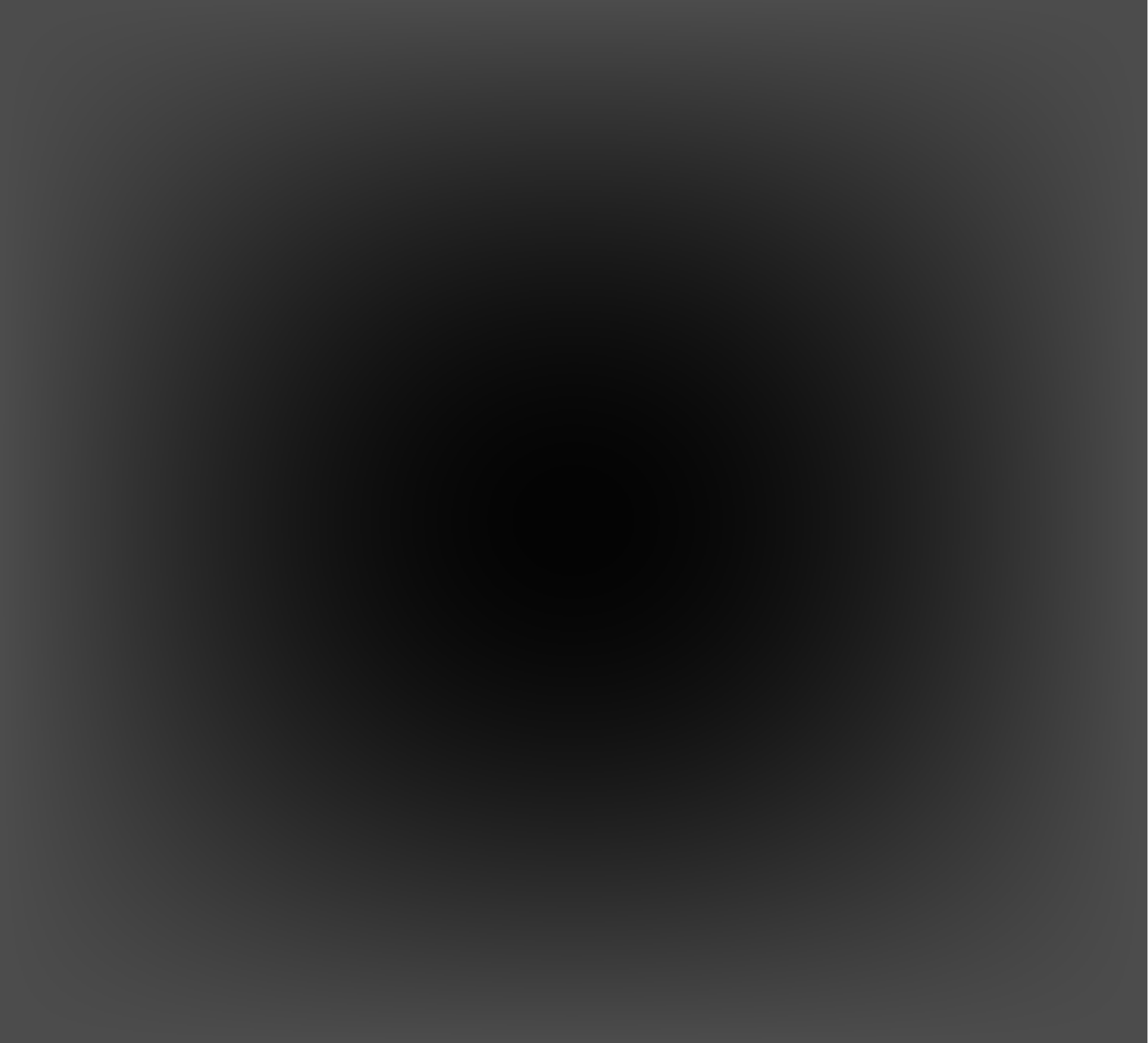}\,%
  \includegraphics[width=0.24\textwidth]{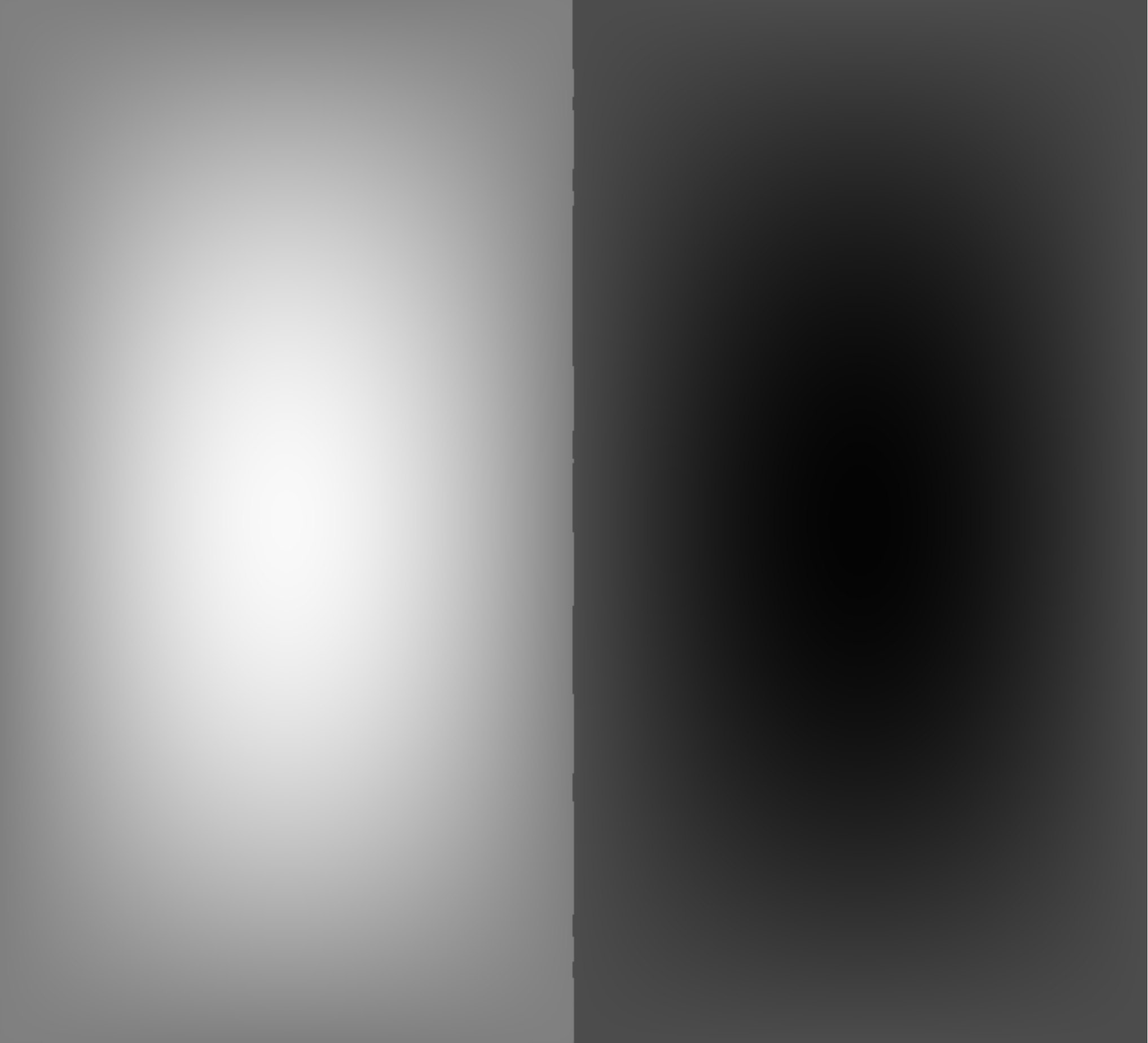}\,%
  \includegraphics[width=0.24\textwidth]{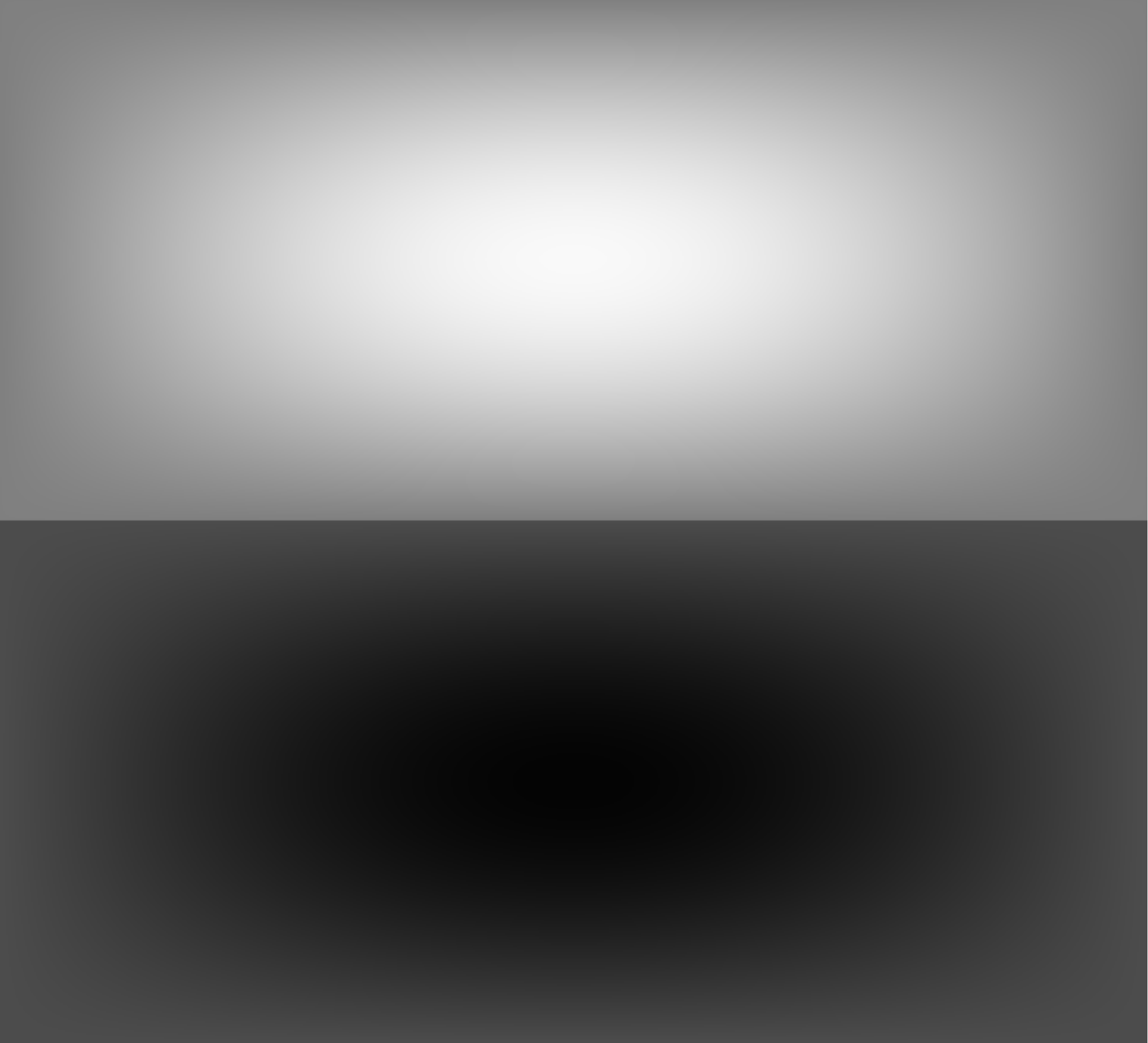}\\
  \includegraphics[width=0.24\textwidth]{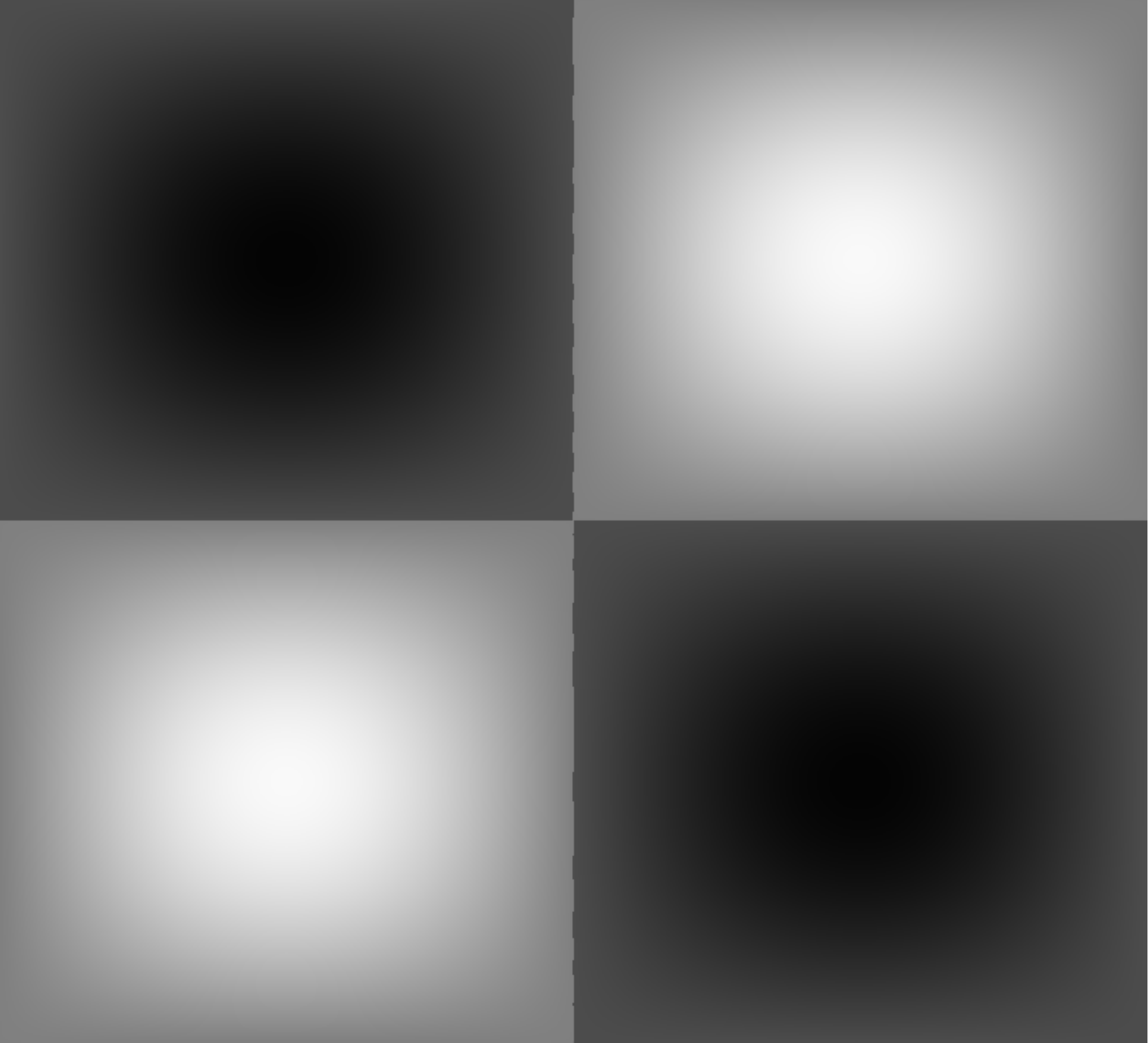}\,%
  \includegraphics[width=0.24\textwidth]{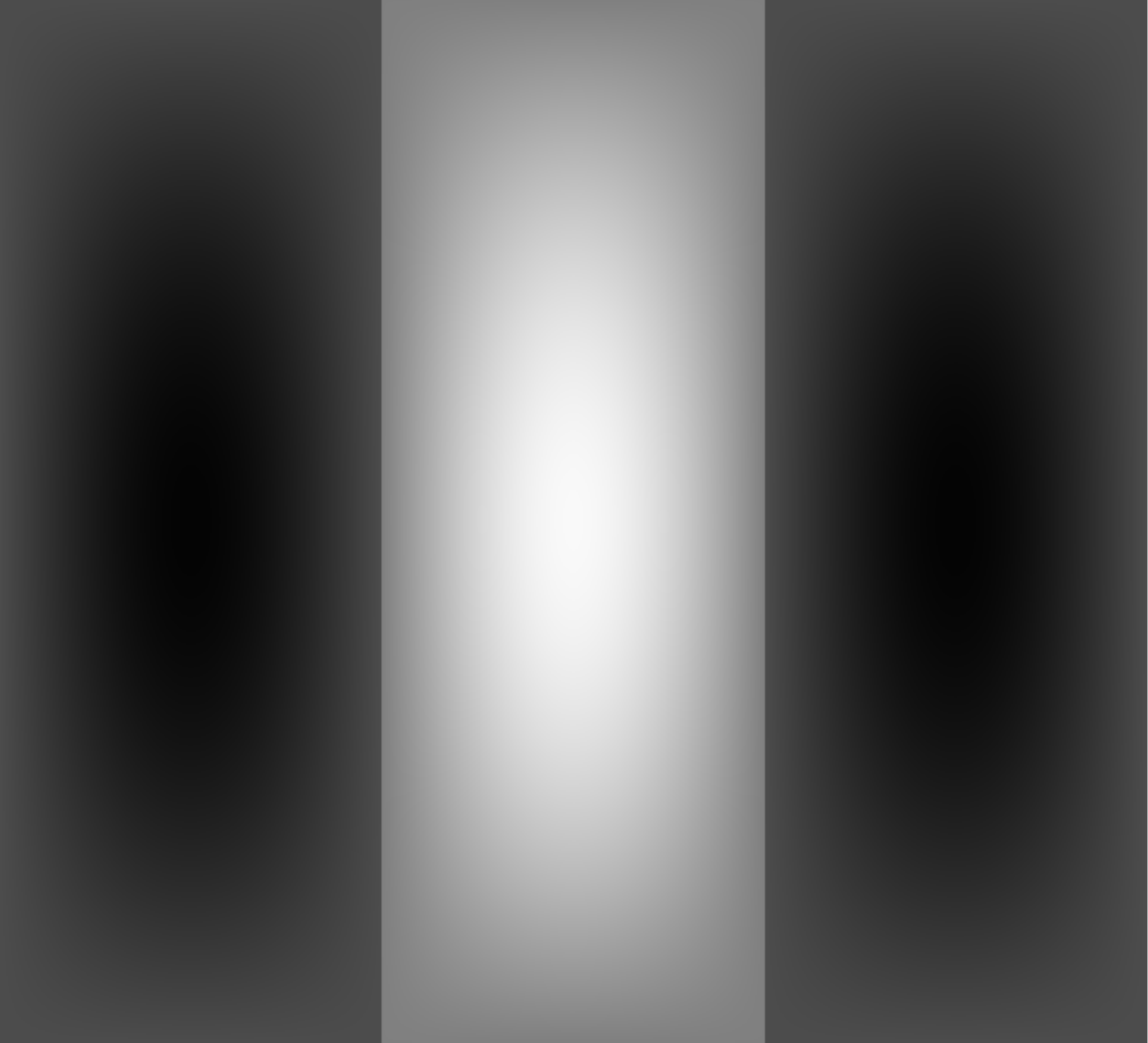}\,%
  \includegraphics[width=0.24\textwidth]{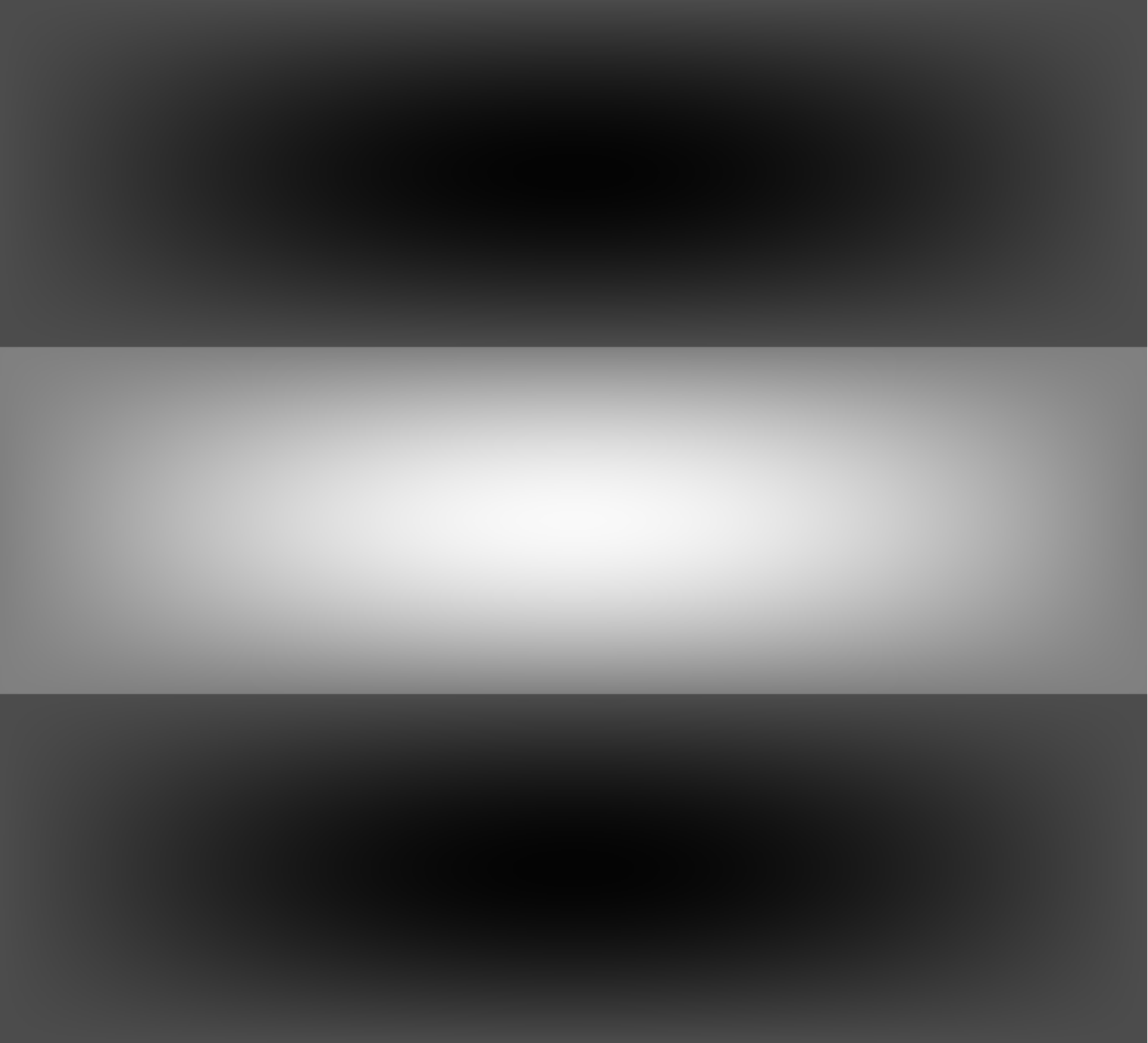}\,%
  \includegraphics[width=0.24\textwidth]{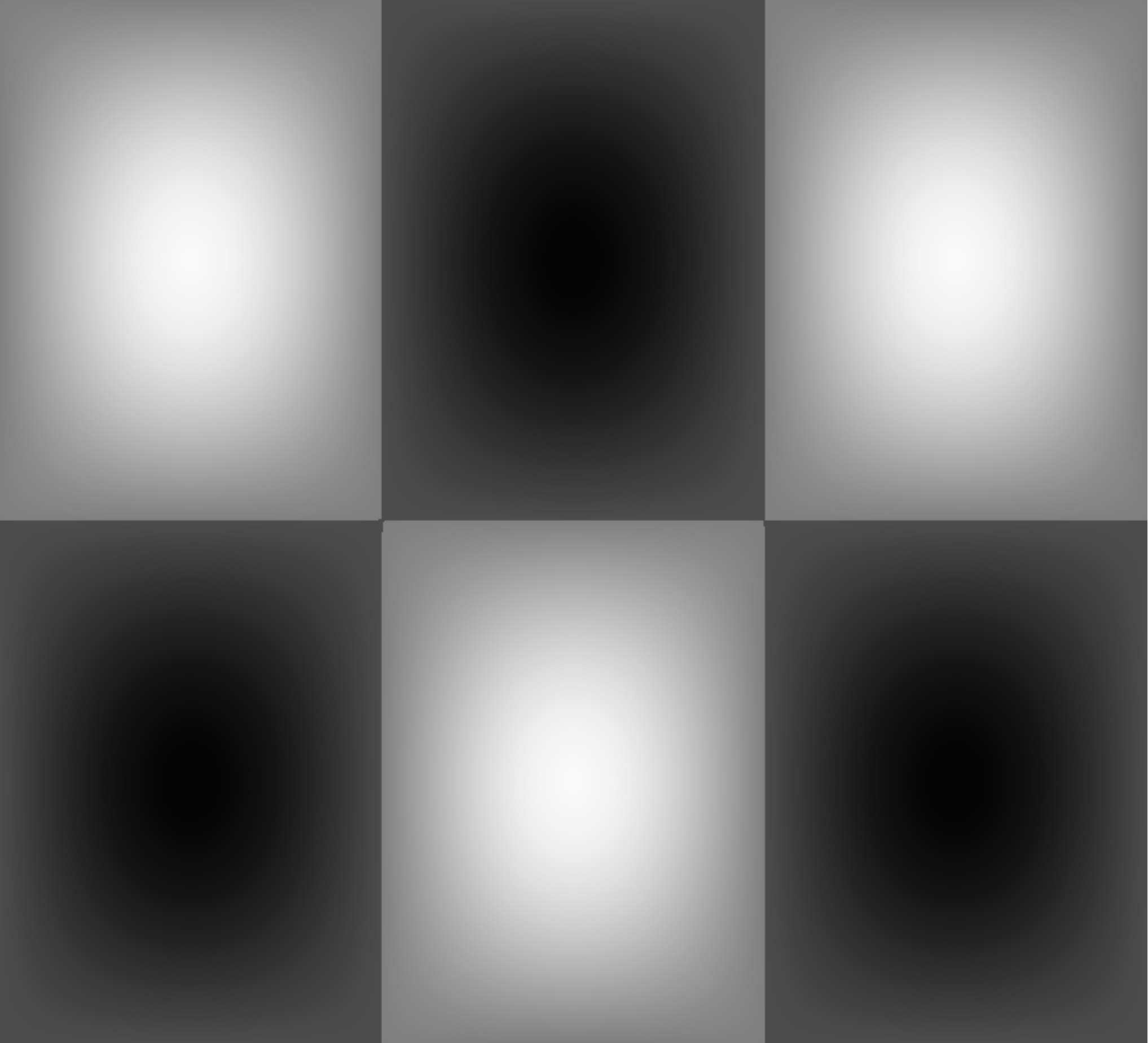}\\
  \includegraphics[width=0.24\textwidth]{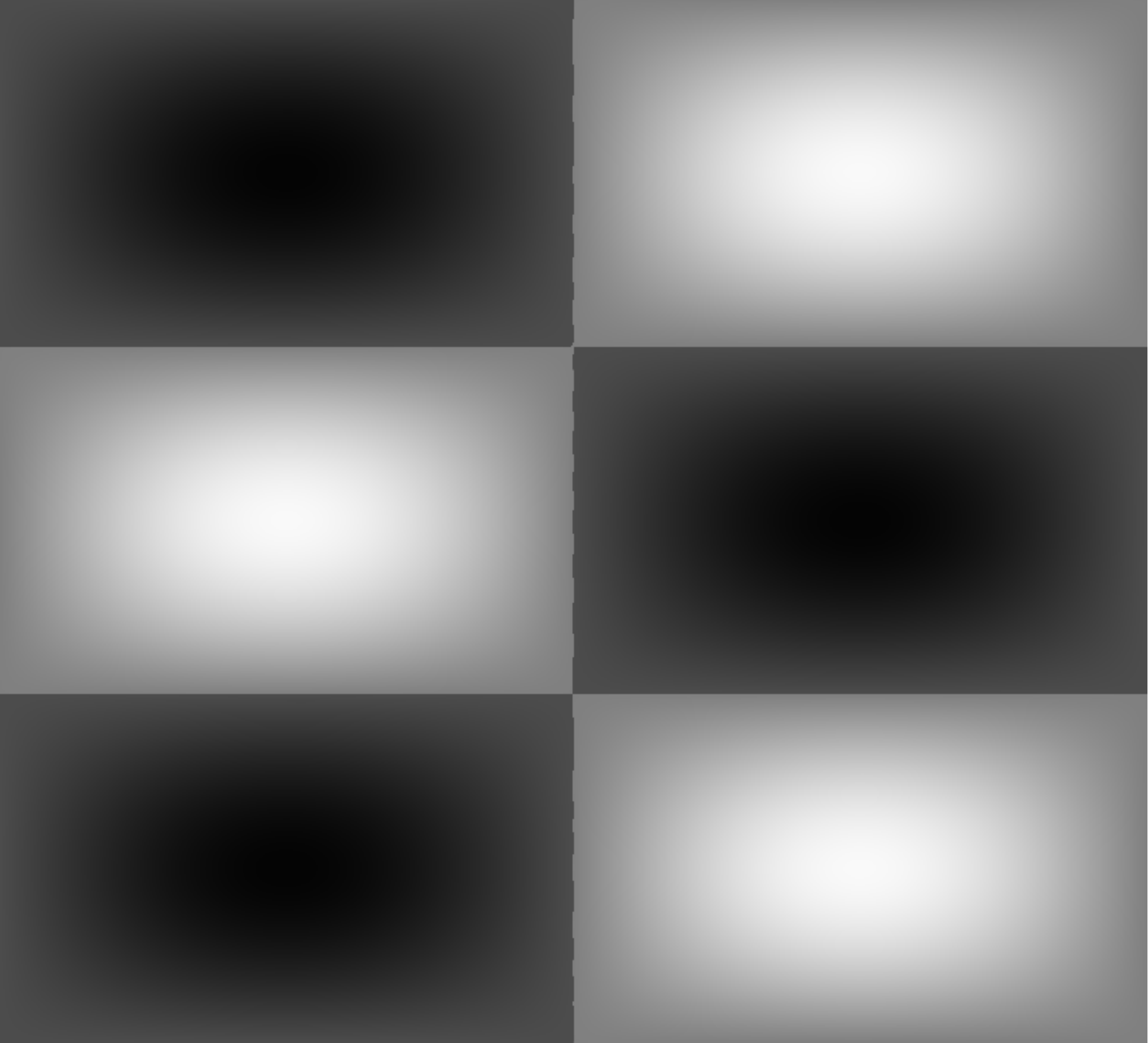}\,%
  \includegraphics[width=0.24\textwidth]{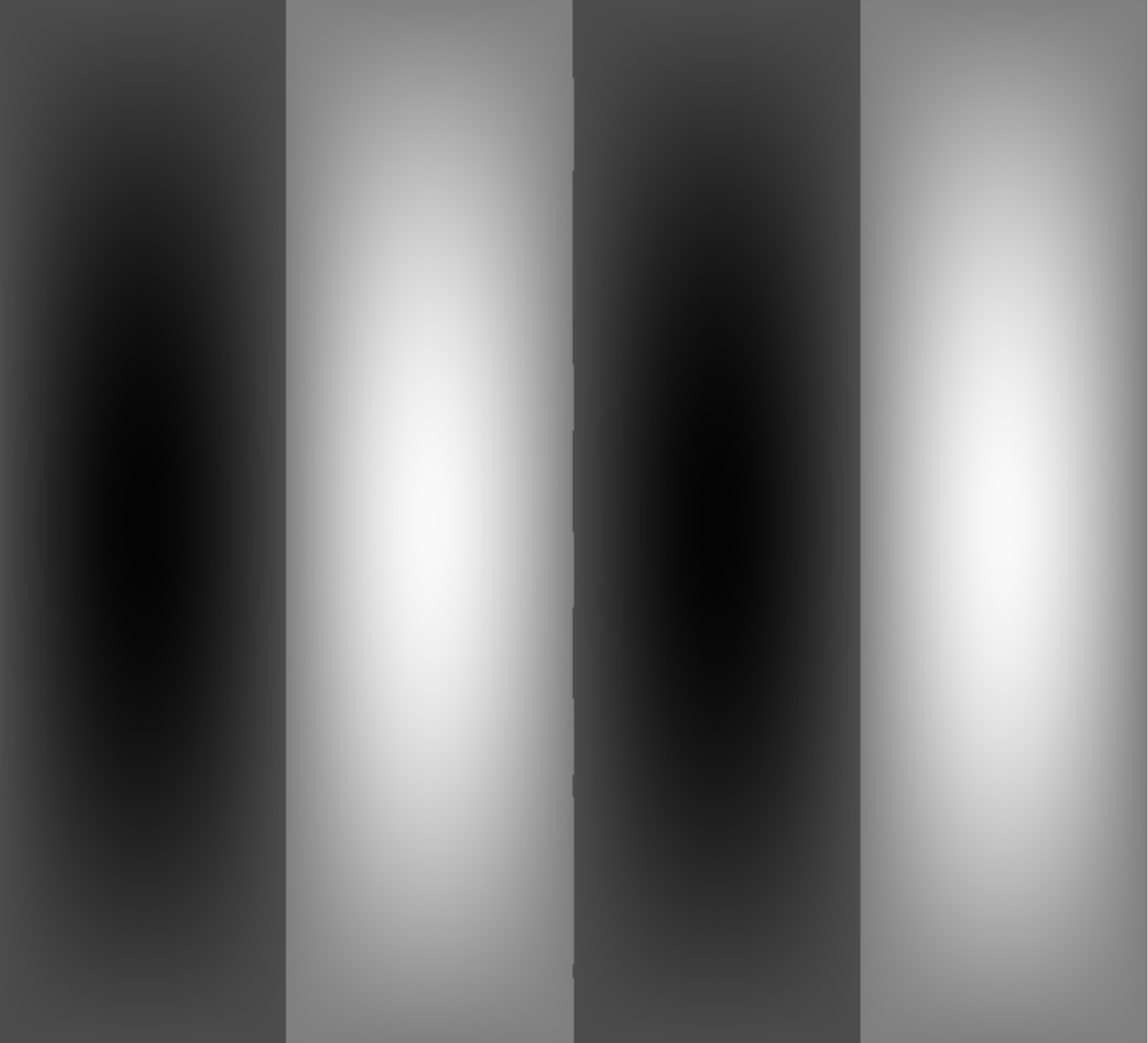}\,%
  \includegraphics[width=0.24\textwidth]{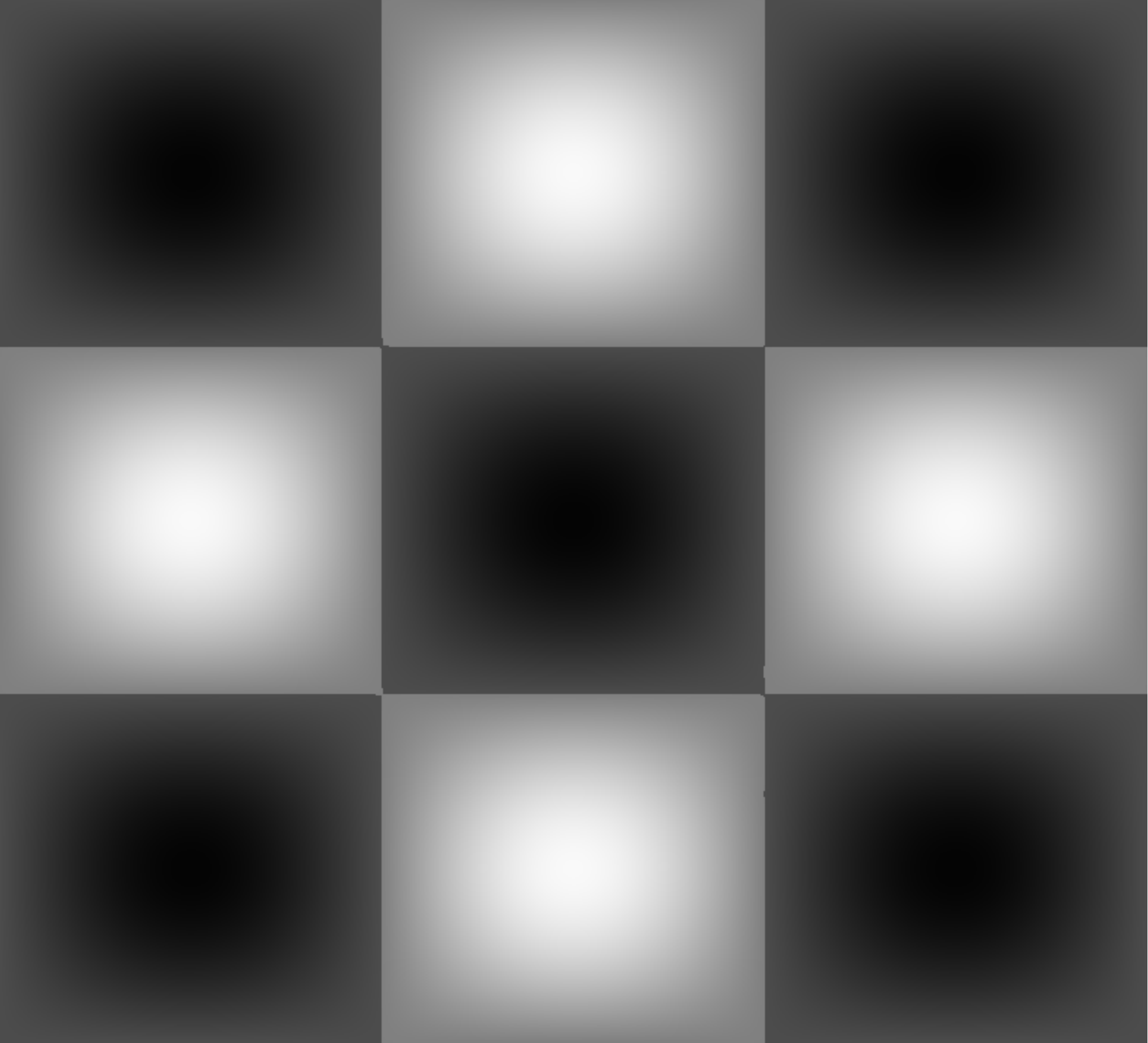}\,%
  \includegraphics[width=0.24\textwidth]{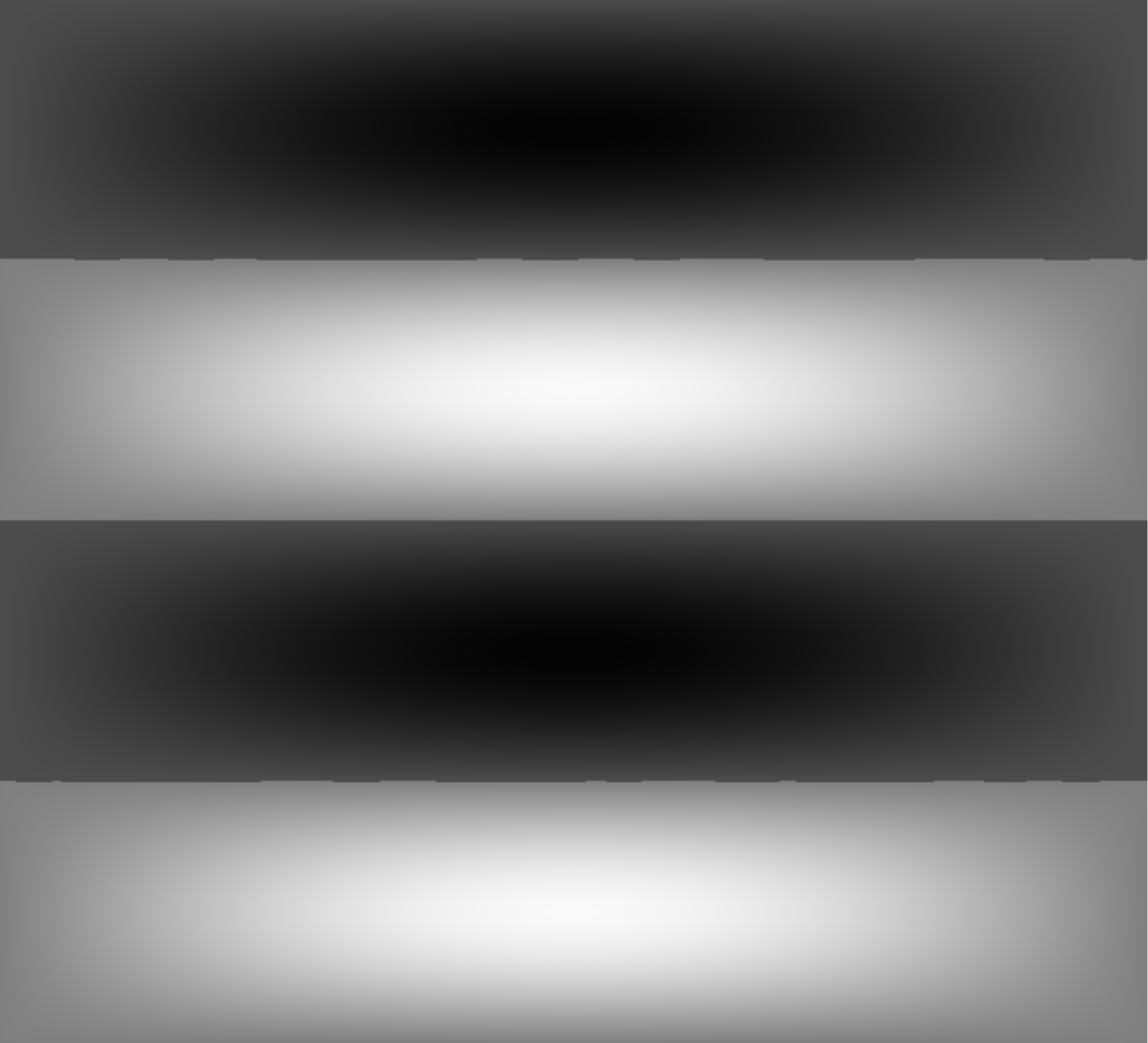}
  \caption{The mesh for the rectangular membrane (upper left image, edge lengths $a=2.2$, $b=2$) consists of $2876$ vertices making up $1395$ triangles with quadratic ansatz. It was generated with parameters $maxArea=0.005$ and $minAngle=30$. The eigenvalues read: $\lambda=\{4.5066,10.6242,11.9089,18.0267,20.8206,24.2467,28.2237,30.3652,35.0971,40.5636,$ $41.5222\}$ (upper left to lower right). The corresponing exact values read $\lambda_e=\{4.5066,$ $10.6241,11.9088,18.0263,20.8200,24.2458,28.2222,30.3633,35.0942,40.5592,41.5176\}$.}
  \label{fig:rectImages}
\end{figure}

Figure~\ref{fig:rectImages} shows a rectangular membrane with edge lengths $a=2.2$, $b=2$. Because of the slight asymmetry, the eigendmodes are no longer degenerated and can be resolved by the FE method quite accurately at least for the lower frequencies. 

In the following, we give a few suggestions for possible exercises that could be done with \emph{NumChladni}.

\begin{itemize}
  \item Given a rectangular membrane with edge lengths $a$ and $b$, $a\neq b$. The edges are fixed. Find the relation between these edge lengths and the eigenfrequencies (see \eref{appeq:efrect}).
  \item Given a rectangular membrane with edge lengths $a$ and $b$, $a\neq b$. The edges are free, and only one point of the membrane is fixed. Describe the differences to the previous example. 
  \item A famous question posed already by Kac~\cite{Kac1966} in 1966 is, ``Can one hear the shape of a drum?'' Find different polygonal shapes that have the same eigenvalues. An example is given by Driscoll~\cite{Driscoll1997}, see also Fig.~\ref{fig:ncDrum}.
  \item Derive the stiffness and mass element matrices in case of a quadratic ansatz function $u_i(x,y)=\upsilon_{(1)i}+\upsilon_{(2)i}x+\upsilon_{(3)i}y+\upsilon_{(4)i}x^2+\upsilon_{(5)i}xy+\upsilon_{(2)i}y^2$.
\end{itemize}

\section{Summary}
In this work the eigenvalue system to find the eigenmodes of arbitrarily shaped, thin membranes by means of the finite element method was derived. The more realistic but also more complicated Kirchhoff plates are defered to future work. 

For an interactive and comfortable exploration of such membranes, the graphical user interface \emph{NumChladni} was developed based on the Qt framework and the Open Graphics Library. Basic problems like the meshing of the membranes and the solution of the eigenvalue system were delegated to specified libraries.
As the source code of \emph{NumChladni} is freely available, the interested student is encouraged to have a look at the coding details. 

\appendix

\section{Rectangular membrane}\label{appsec:quadMembrane}
The rectangular membrane where all four boundary lines are fixed yields the most simple two-dimensional solution of the wave equation \eref{eq:2dwave}. 
For a rectangle with edge lengths $a$ and $b$, these boundary conditions are given by $\psi(t,0,y)=\psi(t,a,y)=\psi(t,x,0)=\psi(t,x,b)=0$ with $x\in[0,a]$ and $y\in[0,b]$.
As we are interested only in the eigenmodes, we use the simplified product ansatz
\begin{equation}
  \psi(t,x,y) = \sin(kx)\sin(\hat{k}y)\cos(\omega t),
\end{equation}
which fulfills the boundary conditions if and only if the wave numbers $k$ and $\hat{k}$ read
\begin{equation}
  k = \frac{r\pi}{a},\qquad \hat{k} = \frac{s\pi}{b},\qquad\mbox{with}\quad r,s=\left\{1,2,\ldots,\right\}.
\end{equation}
Then, we immediately obtain the eigenfrequencies
\begin{equation}
  \omega_{rs}^2 = c^2(k^2+\hat{k}^2) = c^2\pi^2\left(\frac{r^2}{a^2}+\frac{s^2}{b^2}\right)
  \label{appeq:efrect}
\end{equation}
to the corresponding eigenmodes $\psi_{rs}(t,x,y)=\sin(k_rx)\sin(k_sy)\cos(\omega_{rs}t)$.
The nodal lines of the fundamental eigenmodes $\psi_{rs}$ are parallel to either the $x-$ or the $y-$ axis.

If the edge lengths are equal, the eigenfrequencies are degenerated, e.g. $\omega_{23}=\omega_{32}$. Then, different eigenmodes to the same eigenfrequency can be mixed which produces `non-natural' patterns like in Fig.~\ref{fig:quadImages}.



\section*{References}
\bibliographystyle{unsrt}
\bibliography{lit_chladni}
\end{document}